\documentclass[preprint,showpacs,byrevtex]{revtex4}
\usepackage[dvips]{graphicx}
\usepackage{dcolumn}
\usepackage{epsfig}

\newcommand {\tom}{{\tilde \omega}}
\newcommand {\tOm}{{\tilde \Omega}}

\newcommand {\ts} {{\tilde \sigma}}

\begin{document}

\preprint{LAUR-00-5153}
\title[Chiral Phase Transition at Finite Density]
{Equilibrium and  non-equilibrium properties associated with
the chiral phase transition at finite density
in the Gross-Neveu Model}

\author{Alan Chodos}
\email{chodos@aps.org}
\affiliation{American Physical Society,
One Physics Ellipse, College Park, MD 20740}

\author{Fred Cooper}
\email{cooper@schwinger.lanl.gov}
\affiliation{Theoretical Division,\\
   Los Alamos National Laboratory, Los Alamos, NM 87545}
\author{Wenjin Mao}
\email{maow@physics.rutgers.edu}
\affiliation{Department of Physics, Boston College, Chestnut Hill, MA
02167}

\author{Anupam Singh}
\email{singh@lanl.gov}
\affiliation{ Los Alamos National Laboratory, Los Alamos, NM 87545}

\begin{abstract} We study the dynamics of the chiral phase transition at
finite density in
the Gross-Neveu (GN) model in the leading order in large-$N$ approximation.
The phase structure of the GN model in this approximation has the property
that there is a tricritical point at a fixed temperature and chemical
potential
separating regions where the chiral transition is first order from that
where
it is second order. We consider evolutions starting in local thermal and
chemical equilibrium
in the massless unbroken phase for conditions pertaining to traversing
a first or second order phase transition.   We assume boost invariant
kinematics and determine the evolution of
the order parameter $\sigma$, the energy density and pressure  as well as
the
effective temperature, chemical potential and interpolating number densities
as a function of the proper time $\tau$.   We find that before the
 phase transition, the system behaves as
if
it were an ideal fluid in local thermal equilibrium with equation of state
$p=\epsilon$. After the phase transition, the system quickly reaches its
true
broken symmetry vacuum value for the fermion mass and for the energy
density.
The  single particle
distribution functions  for Fermions and anti-Fermions go far out
of equilibrium as soon as the plasma traverses the chiral phase transition.
   We
have also determined the spatial dependence of the ``pion"  Green's function
$
\langle \bar{\psi}(x)  \gamma_5 \psi(x)  \bar{\psi}(0)  \gamma_5 \psi(0)
\rangle$  as a function of the proper time.

\end{abstract}
\maketitle

\section{Introduction}

The phase structure of QCD at non-zero  temperature and baryon density
is important for the physics of neutron stars and heavy ion collisions.
The approximate phase structure for QCD with different numbers of quark
flavors
has been mapped out in various mean field and perturbative
approximations \cite{ref:super1} \cite{ref:super2} \cite{ref:super3}
\cite{ref:ref1}.
The phase structure for two masless quark flavors (up and down) already
reveals a rich
sturcture.  In addition
to the well known chiral symmetry broken
and restored phases,
recent investigations have revealed the possibility of a color
superconducting phase at low temperatures and relatively high densities.
The transition to the superconducting phase as we increase $\mu$ at zero
temperature is first order. On the other hand,
in the chiral condensation regime at zero chemical potential, the
phase transition as we increase the temperature to the unbroken mode is
second order.
This suggests that there is a regime at intermediate chemical potentials
where the chiral phase
transition is first order. Along the  line separating
the broken and unbroken chiral phases there is a tricritical point.

   One of the most pressing experimental questions is to what extent
experiments at the
relativistic heavy ion collider RHIC can
explore this rich phase structure and what would be the experimental
consequences of having
a quark-gluon plasma rather than a hadronic plasma following a collision of
heavy ions. Since the
production and evolution of the quark-gluon plasma in a heavy ion collision
might be a
nonequilibrium process, one  needs to understand the evolution of an
expanding,
possibly out of equilibrium,
plasma.  We have considered a toy model, which has several properties in
common with
two flavor massless QCD to explore these nonequilibrium evolutions.  The
model we have found
\cite{ref:us} is a $1+1$ dimensional model of self interacting fermions,
that has, in the leading order
in large-N (LOLN) approximation a  phase diagram with properties similar to
that of massless two flavor QCD such as
a tricritical point as well as a superconducting phase transition as one
increases $\mu$ at
low temperature.  Since an ultrarelativistic collision leads to an
essentially one dimensional
expansion at early times, it is hoped that the rate of expansion in our toy
model will be similar
to that found in QCD so that the rate the system undergoes the phase
transition will be
similar to what would be found in a a more realistic $3+1$ dimensional
expansion.  Since this model has asymptotic freedom,
the coupling constants run logarithmically in LOLN which is a feature shared
with
 QCD. In this paper we will confine ourselves for studying the dynamics near
the tricritical point in our toy model which for that case reduces to the 
Gross-Neveu model.  

 One of the  questions important for RHIC is
whether there are
unambiguous experimental
signatures resulting from a change in the nature of the phase transition as
a function of
the chemical  potential (baryon density).  In our toy model, the change is
the
difference from a first order to second order transition.  In actuality this
change might
be the change from a first order transition to a crossover phenomena.  Our
approach is to
directly determine the time evolution of the plasma starting from an initial
point on the
phase diagram above the chiral phase transition and watch the evolution
through the
phase transition.  From the time evolving fermion mode functions one can
calculate
many physical quantities such as the current current  correlation function
which determines
the dilepton rate as well as various  particle correlation functions such as
that for the
pion.  In this paper we determine single particle
 distributions
functions as well as the composite particle pion correlation function to
try to find the difference in experimentally measurable quantities when a
plasma
evolutions traversing say a first rather than a second order phase
transition.
For the purpose of studying the chiral phase transition, we can restrict
ourselves
to just a sector of our toy model in which it reduces to   the well known
Gross-Neveu (GN)
model \cite{ref:GN}. This simpler model allows us to study evolutions on
both sides of a tricritical point. The exact phase structure of the 
Gross-Neveu model in dimensions $2 \le 4$ at finite temperature and chemical potential 
has been the subject of several recent investigations \cite{phase}. Using
both renormalization group methods, dimensional reduction methods as well as
strong coupling expansions, it is thought that the line of chiral phase transitions
in all these dimensions is either second order or weakly first order, which is
the same situation as pertains in the leading order in large-N calculation.  
What is missing in the leading order large-N is real scatterings that
could lead to rethermalization. Therefore, the findings of our simulations
that the distributions of fermions and antifermions goes far out of equilibrium
following the transition, might easily be modified by a more realistic simulation.
The calculations presented here must be thought of as presenting the first field theory
simulations at finite chemical potential of an evolution through a chiral phase transition
with a realistic expansion rate for the plasma appropriate to a heavy ion collision.
Future studies will remedy some of the shortcomings of this toy model, in that
inhomogeneous plasmas will be studied as well as $1/N$ resummation methods will 
be used in future simulations which will still be based, however on Gross-Neveu
like models albeit in $3+1$ dimensions as well as in $1+1$ dimensions.  The 
approach we take here is to directly solve the evolution equations of 
a quantum plasma in leading order in large-N.  A complementary approach
is to study critical slowing near the critical point using
ideas from universality and dynamical critical phenomena \cite{crit}.
Our interest is more in having a complete space time picture of an evolving
quark plasma and our hope is that once we resum the LOLN approximation using
Dyson equations and consider the $3+1$ dimensional version of this model that
we will be able to address issues of critical slowing down.  The calculations
presented here in the toy model already exhibit the effects of 
how having small rather than zero quark masses at high temperatures change the
time period of the transition. They also demonstrate how one can calculate
all spatial correlation functions  as well as the time evolution of the
Temperature and chemical potential, and how a hydrodynamic approach can 
be quite accurate before the phase transiton.

   Following a relativistic heavy ion collision, the ensuing plasma expands
and cools traversing
the chiral phase transition.  In hydrodynamic simulations of these
collisions
\cite{Landau,CFS,Bjorken,PRD}, as well as in parton cascade models and
other event genarator approaches
 \cite{Feynman,BjCC,CKS,Lund,Low,Nussinov}, one
finds that it is a reasonable approximation to treat the initial phase of
the expansion
as a 1+1 dimensional boost invariant   expansion
along  the beam ($z$) axis.
In this approximation, the  fluid velocity scales as $z/t$. In
terms of the variables fluid rapidity $\eta = {1 \over 2} ~{\rm ln
}~\left({t+z
\over t-z} \right)$ and fluid proper time  $\tau = (t^2-z^2) ^{1/2}$,
physical quantities such as $\sigma,\epsilon$  become independent of $\eta$,
as discussed in refs.
\cite{CFS,Bjorken,PRD}.  Such an approach was used in our field theory
calculations of
the production and evolution of  disoriented
chiral  condensates in the O(4) $\sigma$ model in ref.\cite{ref:DCC}.
This approximation is valid for particles produced in the central rapidity
region.
To study more peripheral collisions a full inhomogenous calculation must be
performed. This latter study has just started and
 will be the subject of a future paper.

These kinematical considerations translate into the expansion being
homogeneous
in the fluid rapdity $\eta$, which allows us to convert what would be a set
of partial
differential equations for the mode
functions to a  much simpler set of ordinary differential equations in the
parameter $\tau$.
The LOLN approximation we will use in obtaining the field equations has been
 been discussed earlier by ourselves and others in
\cite{PRD,CM,PRL,PRD2,covariant,blsnoneq,cmupittparnoneq,ucsbnoneq}
and applied to the problem of disoriented
chiral  condensates in \cite{ref:DCC,dccbdh}. Extending the boost invariant
simulation to $3+1$ dimensions so that transverse distributions can be studies
is relatively simple. 

In solving the time evolution equations for the quantum fields, the initial
conditions for the
fields are specified at $\tau=\tau_0$, that is, on a hyperbola of constant
proper time.  The $\tau$
evolving energy density and pressure are obtained from the expectation value
of the energy
momentum tensor.  To discuss the production of particles we introduce the
concept of an
adiabatic number operator which is an adiabatic invariant of the LOLN
Hamiltonian.
 Although our equations will be valid for arbitrary initial conditions, to
study the regime
around the tricritical point we will assume that at some initial proper time
the system
can be described by a Fermi-Dirac Distribution with given $\mu_0$ and $T_0$
in the comoving
frame. In our simulations we will also  choose the initial conditions on our
mode functions
to agree with the lowest order WKB approximation result. By choosing this
initial condition
on the mode functions, the
adiabatic number operator then gives a smooth interpolation between the
initial Fermi-Dirac
distribrutions described by
$\mu_0$,
$T_0$ and the  final outstate number operators. The rest of the paper is
organized as follows.
In section II we review the equlibrium properties of the GN model at finite
$\mu$ and $T$ in the
LOLN approximation. Particular attention is paid to the phase diagram.   In
section III we derive
the action in curved coordinates in order  to discuss the evolution in terms
of the parameters
$\tau$ and
$\eta$.   Section IV is concerned with renormalization and obtaining
explicitly finite evolution
equations. In section V we discuss our choice of initial conditions. In
section VI we derive
an expression for the expectation value of the energy momentum tensor and
obtain expressions
for the renormalized energy density and pressure in terms of the mode
functions of the fermion
field.  In section VII we introduce the adiabatic number operator and obtain
simple
expressions for both the fermion and antifermion interpolating number
operators in terms
of the modes.  In section VIII we discuss our numerical results for the
proper time evolution
of the effective fermion mass, $\mu$, $T$ as well as the interpolating
number densities.
In section IX we determine the time evolution of the pion correlation
function.
Some of the results we obtain in this article  were summarized  and
presented at a Riken workshop  \cite{usletter}.

\section{Effective Potential and Phase Structure}
The
Lagrangian  for the Gross-Neveu Model \cite{ref:GN} is
\begin{equation}
 {\cal L} = - i \bar {\Psi}_i \gamma^{\mu}
\partial_{\mu} \Psi_i - {1 \over 2} g^2 \left(\bar {\Psi}_i\Psi^i \right)^2,
\end{equation}
which is invariant under the discrete chiral group: $\Psi_i \rightarrow
\gamma_5 \Psi_i$.
In leading order in large $N$ the effective action is
\begin{equation}
S_{eff}= \int d^2 x \left[ -i
\bar {\Psi}_i \left( \not\partial + \sigma \right)\Psi^i -
\frac {\sigma^2 }{2 g^2 } \right] + {\rm tr}{\rm ln} S^{-1}[\sigma]~,
\label{lolnS}
\end{equation}
where $ S^{-1}(x,y)[\sigma]  = \left(\gamma^{\mu} \partial_{\mu} + \sigma
\right) \delta (x-y)$.

\begin{figure}
   \centering
   \epsfig{figure=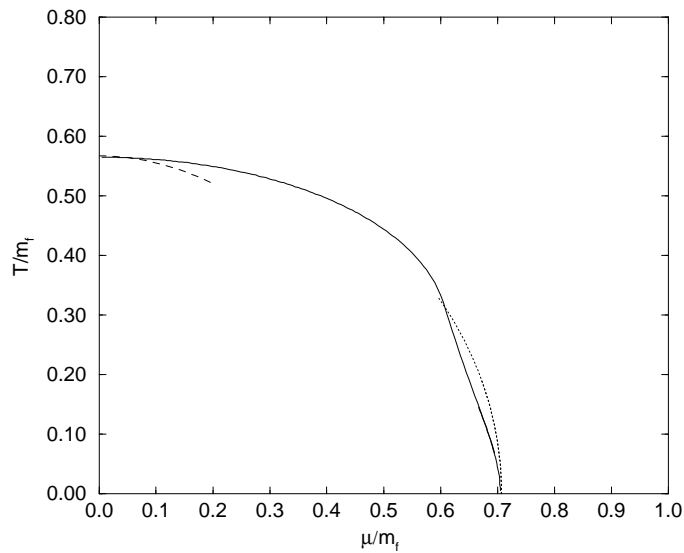,width=4.0in}
\caption{Phase Structure at finite Temperature and Chemical Potential $\mu$. Dotted
lines correspond to approximate analytic results described in the text.}
   \label{fig:phase}
\end{figure}

The phase structure of the GN model at finite temperature
and chemical potential  in this approximation has
been known for a long time \cite{ref:us} \cite{ref:GN2} and is displayed
in Fig. \ref{fig:phase}.

This figure summarizes several facts. In the  GN
model there is spontaneous symmetry breaking of the discrete chiral symmetry
at zero chemical
potential and temperature. The value of the vacuum expectation value of
$\sigma$ at the minimum of
the  effective potential determines and is equal to the mass of the fermion
$m_f$ in this approximation.  At zero temperature, the symmetry is restored
at a critical value of
 the chemical potential  $\mu_c= {m_f \over \sqrt{2}}$. This phase
transition
is first order.  At
zero chemical potential the system undergoes a second order phase transition
to the unbroken symmetry phase as  the temperature is increased. As a result
of these
two facts,  at
some point in the phase
diagram there is a tricritical point which can be determined numerically to
be at
$ {\mu_c \over m_f} = .608,  {T_c  \over m_f} = .318 $. These facts  can be
ascertained by studying  the effective potential in LONL which is given by
\begin{eqnarray}
V_{eff} (\sigma^2, T, \mu) &=& {\sigma^2 \over 4 \pi} [\ln~ {\sigma^2 \over
m_f^2}
- 1] \nonumber \\ &~& \nonumber \\ &-& {2 \over \beta}
\int_0^{\infty} {dk
\over 2
\pi} ~[\ln~ (1 + e
^{- \beta (E - \mu)})+ \ln (1 + e^{- \beta (E  + \mu)})]~.
\end{eqnarray}
The integrals can be evaluated in the high temperature (and small ${\mu
\over T}$) regime.
Keeping the leading terms in the expansion one obtains
\begin{equation}
V_{eff} (\sigma^2, T, \mu)= {\sigma^2 \over 4 \pi} \biggl( \ln ~ {
T^2 \over T_c^2} + {7 \over 2} {\zeta(3) \over \pi^2  T^2}(\mu^2 + {\sigma^2
\over
4}) \biggr) ~,
 \end{equation}
which leads to the relationship:
\begin{equation}
T_c= {m_f \over \pi} \exp [\gamma - {7 \mu^2 \zeta(3) \over 4 \pi^2 T_c^2}]
\end{equation}
for the phase transition temperature in the regime where the transition is
second order
and ${\mu \over T} << 1$.
At small $\mu^2$ one has approximately
\begin{equation}
T_c= {m_f \over \pi} e^{\gamma} [1- {7 \mu^2 \zeta(3) \over 4 \gamma m_f^2
e^{\gamma}}]~.
\end{equation}

In the low temperature regime for the case
$ \sigma \leq \mu$  we can  make an approximation to the Fermi-Dirac
distribution
function that again allows us to perform all the integrals analytically and
determine
an approximate analytic form for the Effective Potential.  We write
the  derivative of the potential in the form:
\begin{equation}
{\partial V \over \partial \sigma} = {\sigma \over 2 \pi} \ln {\sigma^2
\over m_f^2} +
{\sigma \over \pi} \int_0^{\infty} {dk \over E} [2 - \tanh{E+ \mu \over 2T}
- \tanh{E- \mu \over 2T}
]~,
\end{equation}
where $ E = \sqrt{k^2+\sigma^2}$,
and then replace the function  $\tanh({E-\mu \over T})$  with the straight
line interpolation:
\begin{equation}
 \tanh(x) \rightarrow \{ 1 ~{\rm if}~  x>2;~ -1 ~{\rm if}~x < 2 {\rm and}
~ x ~{\rm if}~ |x| \leq 2 \}~.
\end{equation}
Using this approximation the integrals can be performed and $V$ determined.
The results
are shown as the dotted curve in Fig. 1. The analytic expression is given in
\cite{ref:us}.

When $T=0$, the effect of the chemical potential is the most dramatic.
In that limit   $\tanh(x) = \epsilon(x)$, and we obtain the exact result
\begin{eqnarray}
{\partial V \over \partial \sigma}&& = {\sigma \over 2 \pi} \ln {\sigma^2
\over m_f^2} +
{\sigma \over \pi} \int_0^{\sqrt{\mu^2-\sigma^2}} {dk \over E} \Theta(\mu
-\sigma ) \nonumber
\\
 && = {\sigma \over 2 \pi} \ln {\sigma^2 \over m_f^2} +
{\sigma \over \pi} \Theta(\mu -\sigma)  \{ \ln(1+ \sqrt{1-\sigma^2/\mu^2})
-{1 \over 2}
\ln{\sigma^2 \over \mu^2} \} ~. \label{1st1}
\end{eqnarray}
This can be integrated to give the result that for $\sigma \leq \mu$
the effective potential is given by:
\begin{equation}
V_{eff} = {1 \over 4 \pi}  \{  \sigma^2 ( 2 ~ \ln [{\mu + \sqrt{\mu^2 -
\sigma^2}
\over m_f}] - 1) - 2 \mu \sqrt{\mu^2 - \sigma^2} + C(\mu) \} \label{1st2}~,
\end{equation}
wheras, for  $\sigma > \mu$ the effective potential is equal to its $\mu =
0$ value, namely
\begin{equation}
V_{eff} ={\sigma^2 \over 4 \pi} [\ln~ {\sigma^2 \over m_f^2}
- 1] + C(\mu) ~. \label{1st3}
\end{equation}
The integration constant can be fixed by choosing $V_{eff}(\sigma=0)
=0,$ which yields
\begin{equation}
 C(\mu) = {\mu^2 \over 2 \pi}~.
\end{equation}
At $T=0$ in the broken symmetry phase the effective mass is independent of
$\mu$ and is given by $m_f$, its value when $\mu =0, T=0$.
When $\mu^2 > m_f^2 /2$ then the true minimum is at $\sigma =0$.  The
transition at  $\mu^2 = m_f^2 /2$ is a first order transition as can be
determined
by eq. (\ref{1st1}) and eq.(\ref{1st2}).  In the toy model \cite{ref:us}
with two
coupling
constants, which also has a superconducting phase, the first
order transition takes
place at the point:
\begin{equation}
\mu^2 = { m_f^2 \over 2}(1- e^{-4 \pi \delta}) ~,
\end{equation}
where $\delta$ is the difference of the inverse of the two coupling
constants of the
model \cite{ref:us}, namely $\delta = {1 \over \kappa} - {1 \over 2
\lambda}$.
When the second coupling constant $\kappa \rightarrow 0^+$, the toy model
reduces
to the GN model.

\begin{figure}
   \centering
    \epsfig{figure=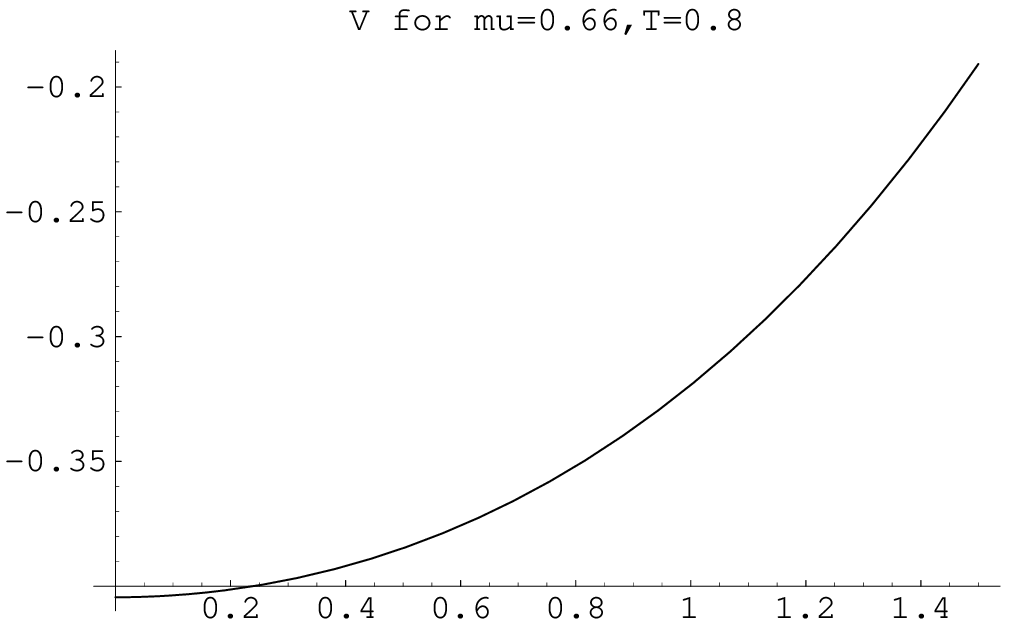,width=2.5in,height=2.0in}
    \epsfig{figure=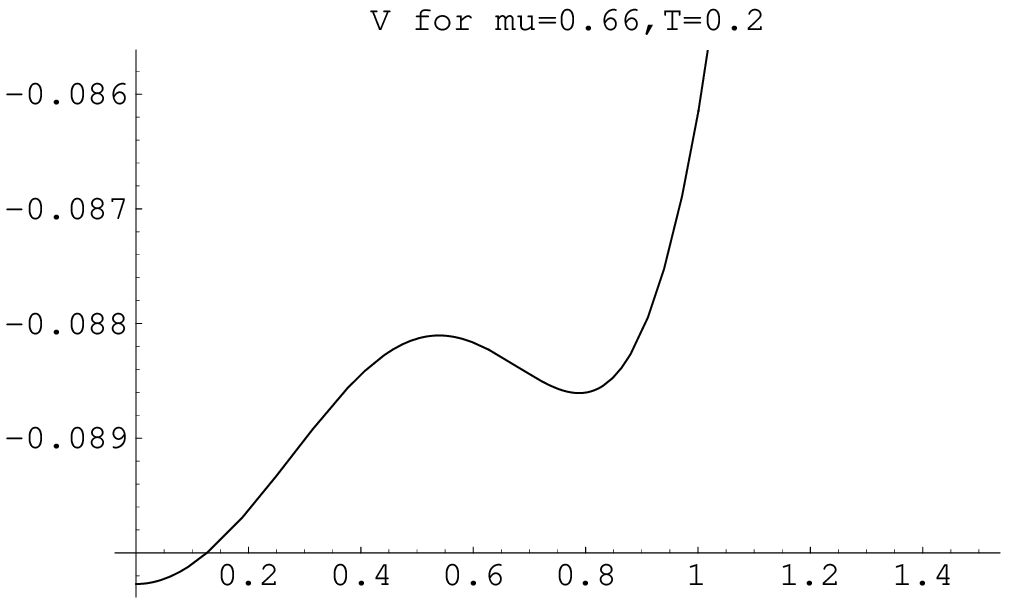,width=2.5in,height=2.0in}
    \epsfig{figure=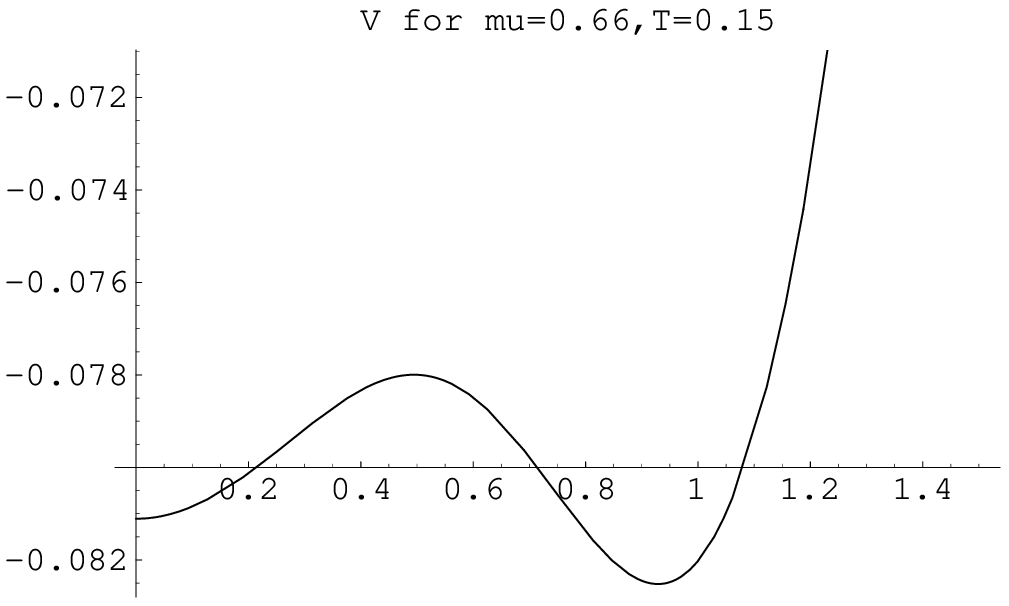,width=2.5in,height=2.0in}
    \epsfig{figure=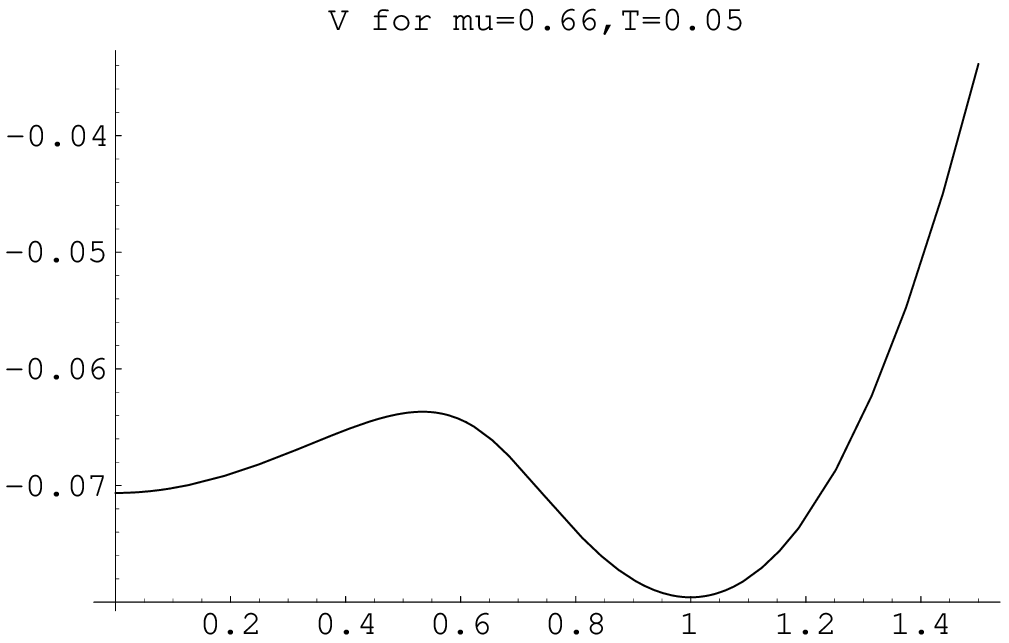,width=2.5in,height=2.0in}
\caption{Evolution of $V_{eff}$ as a function of $T$.
This is for a first order transition.}
\label{Potfirst} \end{figure}

\begin{figure}
   \centering
    \epsfig{figure=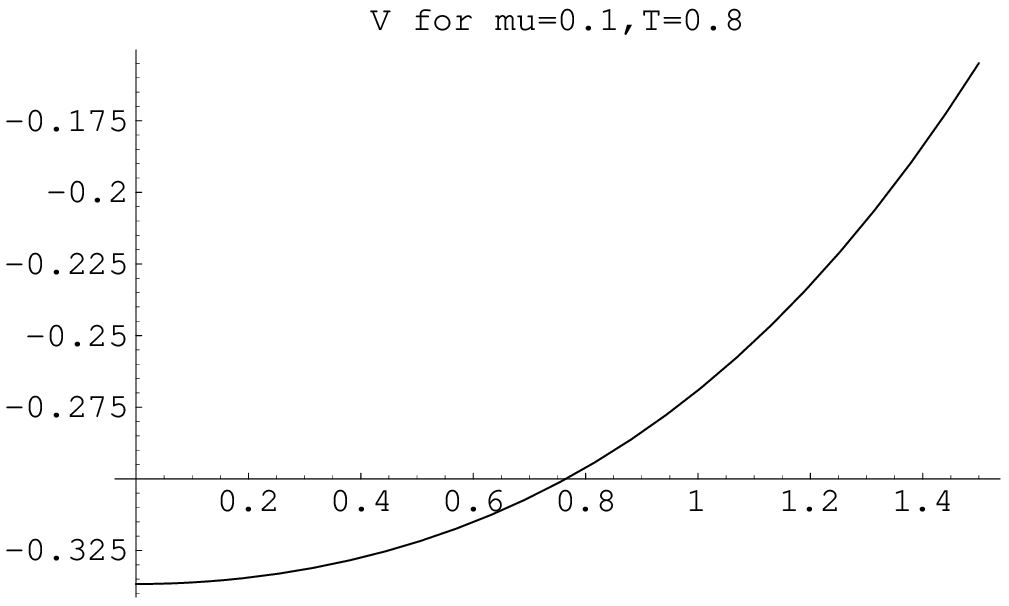,width=2.5in,height=2.0in}
    \epsfig{figure=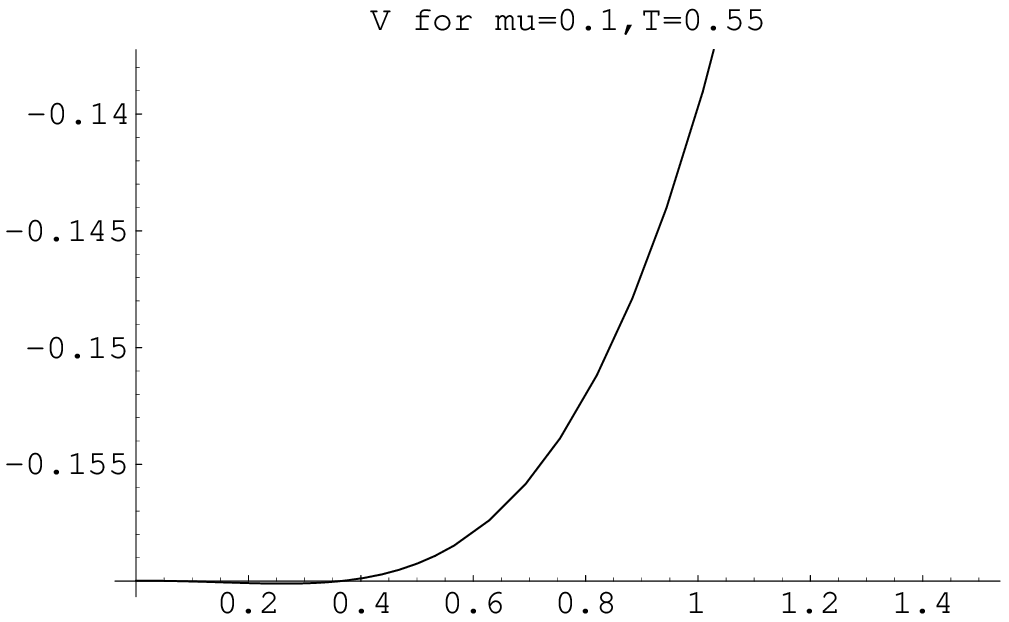,width=2.5in,height=2.0in}
    \epsfig{figure=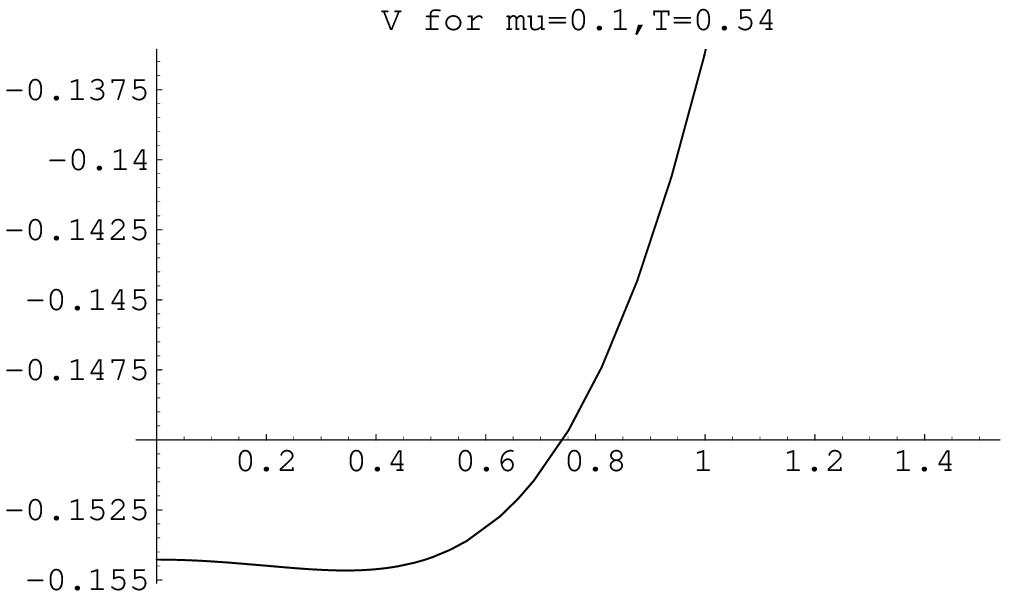,width=2.5in,height=2.0in}
    \epsfig{figure=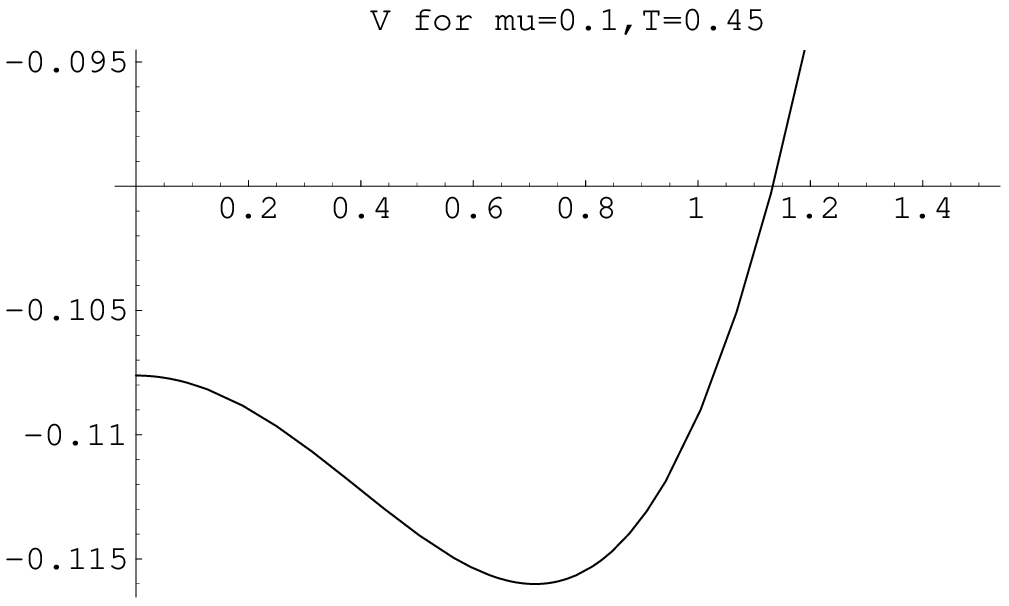,width=2.5in,height=2.0in}
\caption{Evolution of $V_{eff}$ as a function of $T$.
This is for a second order transition.}
\label{Potsecond} \end{figure}

In our straight line interpolation of the $\tanh$ function we obtain for the
tricritical point
which seperates the regime between the first and second order phase
transitions: $
{\mu_c \over m_f} = .661, ~~~  {T_c  \over m_f} = .31 $
as opposed to the exact result
\begin{equation}
 {\mu_c \over m_f} = .608  , ~~~
~~~  {T_c  \over m_f} = .318 .
\end{equation}
In Fig. \ref{fig:phase}  we plot the exact numerical result for the phase
diagram along with  these two approximate results. In Figs. \ref{Potfirst},
\ref{Potsecond} we show the evolution of the
effective
potentials for the first order and second order phase transitions when we
keep
$\mu$ fixed on two sides of the critical value and we decrease the
temperature.

\section {Gross-Neveu model  in curvilinear coordinates}

In order to  make best use of the kinematic constraint that we are
in a scaling regime where the fluid velocity
is  $v=z/t$, we make a coordinate transformation to the light-cone variables
$\tau$ and $\eta$, which are
the fluid proper time and rapidity respectively.
These coordinates are defined
in terms of the ordinary lab-frame Minkowski time $t$ and coordinate along
the beam
direction $z$ by
\begin{equation}
       z= \tau \sinh \eta  \quad ,\quad t= \tau \cosh \eta \,.
\label{boost_tz.tau.eta}
\end{equation}
We shall use the metric convention $(-+)$ which is commonly used in the
curved-space
literature. In what follows, we use Greek indices for the curvilinear
 coordinates  $\tau$ and $\eta$, and Latin
indices for the Minkowski coordinates $z$ and $t$. To obtain the Fermion
evolution equations in the new coordinate
system it is simplest to use a coordinate covariant action such as that
used in field theory in curved spaces, even though here the curvature is
zero.

 The Minkowski line element in these coordinates has the form
\begin{equation}
{ds^2} = {- d \tau^2 + {\tau}^2 {d \eta}^2 }\,.
\label{boost_line_element}
\end{equation}
Hence the metric tensor is given by
\begin{equation}
 g_{\mu \nu} = {\rm diag} (-1, \tau^2)~.
\label{boost_metric}
\end{equation}
with its inverse determined from $g^{\mu \nu} g_{\nu \rho} =
\delta^{\mu}_{\rho}$.
This metric is a special case of the Kasner metric
\cite{Birrell_Davies}.

 The vierbein $V^a_{\mu}$  transforms the
curvilinear coordinates to Minkowski coordinates,
\begin{equation}
g_{\mu \nu} = V^{a}_{\mu}  V^{b}_{\nu} \eta_{ab}\,,
\label{boost_gmunuD}
\end{equation}
where
$\eta_{ab}= {\rm diag} \{-1,1 \}$\@ is the flat Minkowski metric.
A convenient choice of the vierbein for the metric (\ref{boost_metric})
for our
problem is
\begin{equation}
   V_{\mu}^{a}  = {\rm diag}\{1,\tau \} ~,
\end{equation}
so that
\begin{equation}
V^\mu_a= {\rm diag}\left\{1,{\frac {1}{\tau}}\right\}\,.
\label{boost_vierbein}
\end{equation}
The determinant of the metric tensor is given by
\begin{equation}
\det V = \sqrt{-g} = \tau\,.
\label{boost_detV}
\end{equation}

The action for the Gross-Neveu Model in general curvilinear coordinates
(see \cite{Birrell_Davies}) with metric
$g_{\mu \nu}$ is
\begin{equation}
S= \int d^2 x\, ({\rm{det}}\, V) \left[ {\frac{-i}{2}}
\bar {\Psi}_i \tilde{\gamma}^{\mu}
\nabla_{\mu} \Psi^i+ {\frac{i}{2}} (\nabla^{\dag}_{\mu}\bar {\Psi}_i )
\tilde{\gamma}^{\mu} \Psi^i  -i  \sigma \bar {\Psi}_i\Psi^i -
\frac {\sigma^2 }{2 g^2 } \right] ~.
\label{boost_Sf}
\end{equation}
The covariant derivative  is (see \cite{Weinberg})
\begin {equation}
\nabla_{\mu} \Psi_i \equiv (\partial_{\mu} + \Gamma_{\mu}) \Psi_i ~,
\end{equation}
where the spin connection $\Gamma_{\mu}$ is given by
\begin {equation}
\Gamma_{\mu}=\frac{1}{2}\Sigma^{ab}V_{a {\nu}}(\partial_{\mu} V_b^{\nu}
+\Gamma^{\nu}_{\mu \lambda} V^{\lambda}_b ) \,,\qquad
\Sigma^{ab}=\frac{1}{4}[\gamma^a,\gamma^b]\,,
\label{boost_nabla}
\end{equation}
with $\Gamma^{\nu}_{\mu \lambda}$ the usual Christoffel symbol.
The label $i = 1 \ldots N$ corresponding to the SU(N) symmetry.
For the metric eq. (\ref{boost_metric})  (see \cite{Parker}) one finds
\begin{eqnarray}
\Gamma_{\tau}&=&\Gamma_x=\Gamma_y=0  ~,\nonumber \\
\Gamma_{\eta}&=&-\frac{1}{2}\gamma^0 \gamma^3~ .
\label{boost_Gamma}
\end{eqnarray}
The coordinate dependent gamma matrices
$\tilde{\gamma}^{\mu}$ are obtained from the usual Dirac gamma matrices
$\gamma^{a}$
via
\begin{equation}
 \tilde{\gamma}^{\mu} = \gamma^{a} V_{a}^{\mu}(x)\ .
\end{equation}
The coordinate independent Dirac matrices $\gamma^{a}$ satisfy the
usual gamma matrix algebra:
\begin{equation}
\{\gamma^{a},\gamma^{b}\} = 2 \eta^{ab}~.
\label{boost_gammaD}
\end{equation}
{}From the action eq. (\ref{boost_Sf}) we obtain the Heisenberg field
equation for the Fermions,
\begin{equation}
\left( \tilde{\gamma}^{\mu}\nabla_{\mu} + \sigma \right) \Psi_i=0\,,
\end{equation}
which takes the form
\begin{equation}
\left[ \gamma^0 \left(\partial_\tau+\frac{1}{2\tau}\right)
+ \frac{\gamma^3}{\tau} \partial_\eta + \sigma  \right] \Psi_i =0 ~.
\label{boost_Dirac}
\end{equation}
Variation of $S$ with respect to $\sigma $ yields the constraint
equation:
\begin{equation}
\sigma  = - i g^2 \bar  \Psi_i  \Psi^i =  - i {\lambda \over N} \bar  \Psi_i
\Psi^i ~,
\end{equation}
which defines the rescaled coupling constant $\lambda$.  Since we will be
interested in having $N$ copies of the Fermion quantum field, the rescaled
coupling constant is the relevant one for discussing the large-$N$ limit.
The lowest order in large $N$ (LOLN) approximation is obtained by
integrating over the
Fermi
degrees of freedom in the generating functional for the Green's function and
keeping the saddle point contribution in the integral over the constraint
field $\sigma$.  One obtains that the gap equation in leading order
is:
\begin{eqnarray}
\sigma &&
= - i {\frac{\lambda}{2 N}}\left \langle \left[ \Psi^{\dag}_i, \tilde
{\gamma}
^{0} \Psi^i  \right]\right \rangle \equiv  - i {\frac{\lambda}{2 }}\left
\langle \left[
\psi^{\dag}, \tilde {\gamma} ^{0} \psi
\right]\right \rangle ~ .
\label{gap}
\end{eqnarray}
where we have assumed there are $N$ identical $\Psi_i = \psi$.
In the scaling regime $v=z/t$,  the order parameter, which
is the effective
Fermion
mass, is independent of $\eta$ and is a function of $\tau$ only.

For the heavy ion collision problem, we
want to solve these equations subject to initial conditions specified
on the hyperboloid
$\tau = \tau_{0}$.  In LOLN, specifying  the
initial value of the density matrix is equivalent to specifying  the initial
particle-number density and anti-particle number density  with respect
to an
adiabatic vacuum state [see  below]. To complete the specification of the
initial state, the mode functions for the Fermi field also need to be
specified
at $\tau = \tau_{0}$.

The Dirac equation reduces to its Minkowski form if we do a rescaling
\begin{equation}
\psi(x)=  {1 \over \sqrt{\tau}} \Phi(x) ~,
\end{equation}
and introduce the  conformal time   $u$ via
\[    m \tau = e^u . \]
One then obtains
\begin{equation}
\left[ \gamma^0 \partial_ u
+ \gamma^3 \partial_\eta + \tilde{\sigma}  (u) \right] \Phi(x)  =0 ~,
\label{boost_Dirac3}
\end{equation}
where
\[ \tilde{\sigma}  (u) = \sigma \tau = {\sigma  \over m}  e^u ~. \]

Our assumption that the evolution is homogeneous in the rapidity variable
$\eta$ allows us to
expand the Fermion field $\Phi$  in terms of Fourier modes in the momentum
$k_\eta$ conjugate to $\eta$ at
fixed conformal time $u$,
\begin{equation}
\Phi (x) = \int {d k_{\eta} \over 2 \pi} \left[b({k})
\phi^{+}_k(u)
 e^{i k_{\eta} \eta}
+d^{\dagger}({{-k}}) \phi^{-}_{{{-k}}}(u)
e^{-i k_{\eta} \eta}   \right]~.
\label{boost_fieldD}
\end{equation}
The $\phi^{\pm}_{{ k}}$ then obey
\begin{eqnarray}
 \left[\gamma^{0} {d\over d u }
+i {\gamma^{3}} k _{\eta}
 + \tilde{\sigma}(u) \right]\phi^{\pm}_{{k}}(u) = 0~.
\label{eq:phiu}
\end{eqnarray}
 The superscript $\pm$ refers to positive- or
negative-energy solutions with respect to the adiabatic vacuum at $\tau=
\tau_{0}$ as we shall show.
It is convenient to square the Dirac equation by letting:
\begin{equation}
\phi^{\pm}_{{ k}}(u) =  \left[-\gamma^0 {d\over d u}
- i {\gamma^{3}} k_{\eta}
 + {\tilde\sigma}(u ) \right] f ^{\pm}_{k}(\tau) \chi^{\pm} ,
\label{boost_mode_eq_f}
\end{equation}
where the momentum independent spinors $\chi^{\pm} $ are chosen to be
the eigenstates of $
i\gamma^0$

\begin{equation}
i \gamma^{0}\chi^{\pm} =  \pm \chi^{\pm} ~,
\end{equation}
and obey the normalization condition:
\begin{equation}
\chi^{\dag}_{r} \chi_{s} =  \delta_{rs} ~,
\label{boost_e.v}
\end{equation}
with  $ r,s = \pm$.
An explicit representation for these spinors is found in the appendix.
The quantities $f ^{+}_{k}(\tau) \chi^{+}$ and $ f ^{-}_{k}(\tau) \chi^{-}$
are the two linearly independent solutions corresponding to the positive and
negative energy solutions.
When $k_{\eta} =0$ we have that (in what follows the $\eta$ in $k_\eta$
 when $k_\eta$ is a subscript is
suppressed for notational simplicity)
\[   \left[ \mp i {d \over du} + \ts \right] f^{\pm}_{k=0} =0 ~,\]
so  that
\begin{equation}
f^{\pm}_{k=0} \propto e^{\mp ~i~\int \ts du}  ~~.
\end{equation}
The mode functions $f^\pm$ obey the second order differential
equation:

\begin{equation}
\left(- \frac{d^2}{d u^2}-
\tom_k^2  \pm  i~~{d \ts \over d u}  \right )
f^{\pm}_{k}(u) = 0~,
\label{eq:mode}
\end{equation}
where now
\begin{equation}
 \tom_k^2= k_{\eta}^2 +\ts^2(u)~ .
\label{boost_omega_D}
\end{equation}
If we impose the canonical anti-commutation relations on the
fields
\[ \{ \phi_{\alpha}(u,\eta),\phi^{\dag}_{\beta}( u , \eta^\prime) \} =
\delta_{\alpha \beta} \delta(\eta-\eta^\prime)~,
\]
and assume the usual canonical anti-commutation
relations on the Fock space mode operators,
\begin{equation}
\{ b_k,b^{\dag}_q  \} = \{ d_k,d^{\dag}_q \}
=2 \pi \delta(k-q)~,
\end{equation}
then we obtain the condition:
\begin{equation}
\phi^{+\dag}_{k \alpha} \phi^{+}_{k \beta} =
\phi^{-\dag}_{k \alpha}
 \phi^{-}_{k \beta}= {1 \over 2} \delta_{\alpha \beta}
 ~. \label{eq:norm0}
\end{equation}
Taking the trace yields the normalization condition:
\begin{equation}
\phi^{+\dag}_{k} \phi^{+}_{k} =
\phi^{-\dag}_{k}
 \phi^{-}_{k}= 1 ~. \, \label{eq:norm}
\end{equation}
Using eq. (\ref{eq:phiu}) one obtains
\begin{equation}
 {d \over du}  [\phi_k^{\dag ~a} \phi_k^b ] =0 ~,
\end{equation}
where $a,b = +,-$, for every $a,b$.
Reexpressing this in terms of the $f^a$ we have for each Fourier mode at
all $u$
 \begin{equation}
\phi^{\pm \dag}_k \phi^{\pm}_k =  \dot{f}^ {\ast \pm}_k \dot{f}^{\pm}_k
+ \tom_k^2  f^{\ast \pm}_k f^{\pm}_k \pm i \ts \left (   \dot{f}_k^{\pm}f^{
\ast \pm}_k -\dot{f}_k^{\ast
\pm}f^{\pm}_k\right) = 1~ .
\label{eq:uga}
\end{equation}

We also have that
\begin{equation}
 \phi^{\dag +}_k \phi^{-}_k = k_{\eta} \left( f^ {\ast + }_k \dot{f}_k^{-}-
\dot{f}_k^ {\ast + }{f}_k^{-} \right)~. \label{rhs} \end{equation}
Since the right hand side of eq. (\ref{rhs}) is proportional to the
Wronskian and is thus
independent of the conformal time, if we initially choose
\begin{equation}
0 =  f_k^ {\ast + } \dot{f}_k^{-}- \dot{f}_k^ {\ast + }{f}_k^{-}~,
 \label{eq:muga}
\end{equation}
then the two solutions $\phi^ {+} , ~~ \phi^{-}$ remain orthogonal at all
times. This relation as well as eq.(\ref{eq:mode}) can be satisfied by
having
\[ f_k ^{\ast + } = f^{-}_k  ~. \]
These results can be summarized by
\begin{equation}
\phi^{\dag a}_k \phi^b_k = \delta^{ab} ~,
\end{equation}
with $a,b$ taking on either $+$ or $-$.
For both renormalization purposes as well as to introduce the concept
of adiabatic number operators, it is useful to also have a WKB-like
parameterization of the positive-energy solutions $f^{+}_{k}$
as discussed in Ref.~\cite{PRD}:
\begin{eqnarray}
f_{k}^+ (u) = N_{k}
 \frac {1}{\sqrt{2\tOm_{k}}} \exp\left \{ \int_{0}^{u}
\left ( -i\tOm_{k} (u')
- \frac {{\dot{\ts}
}(u')}
{2\tOm_{k} (u')}
\right )
du'\right \} ~.
\label{boost_ansatz_D}
\end{eqnarray}
The $u$ derivative is given by
\begin{equation}
\dot{f}_{k}^+ (u) = \left(- i {\tOm_k}- \Delta_k
\right)  f_{k}^+ (u) ~,\label{eq:fdot}
\end{equation}
where
\begin{equation}
\Delta_k = { \dot{\tOm}_k + \dot{\ts} \over
2 \tOm_k} ~. \label{eq:Del}
\end{equation}
The $\tOm_{k}$ obey the real equation
\begin{equation}
 \frac {\ddot\tOm_{k} + \ddot\ts_{k}}{2 \tOm_{k}} -
{(\dot{\ts} + \dot{\tOm}_k)(\dot{\ts} + 3 \dot\tOm_k) \over 4
\tOm_k^2}
=\tom_k^2(u) - \tOm_{k}^2 ~,
\label{boost_WKB_D}
\end{equation}
which is the starting point for a WKB expansion of the mode functions.
$\tOm_k$ and  $\dot{\tOm}_k$ can be determined from the mode functions as
follows:
\begin{equation}
\tOm_k = -{\rm Im}{\dot{f}_k^+ \over f_k^+}; ~~ -2 {\rm Re}{\dot{f}_k^+
\over
f_k^+} ={\dot{\ts} + \dot{\tOm}_k  \over 2 \tOm_k} ~.
\end{equation}
Using the normalization condition  eq. (\ref{eq:uga}) we can show that
\begin{equation}
Z_k(u) \equiv | f_k^{\pm}|^2 =  \left(\tOm_k^2+ \tom_k^2 + 2 \tOm_k
\ts + ({\dot{\ts}+\dot{\tOm}_k
\over 2 \tOm_k})^2 \right)^{-1}.
\label{eq:zu}
\end{equation}
The normalization of the wave function, $N_k$ is time independent and can
be evaluated at $u=0$. Using eq. (\ref{eq:zu}) and eq.
(\ref{boost_ansatz_D})
we obtain
\begin{equation}
N_k^2 = 2 \tOm_k(u=0) Z_k(u=0) \label{eq:Nk1}~.
\end{equation}
We can now obtain an expression for the the gap equation in terms of the
mode
functions.
Using the mode decompositon and the definitions:
\begin{eqnarray}
< b^{\dag}(k) b(q) > &&= 2 \pi \delta(k-q) N_+(q)  ~,\nonumber \\
< d^{\dag}(k) d(q) > &&= 2 \pi \delta(k-q) N_{-}(q)  ~,
\end{eqnarray}
we obtain
\begin{eqnarray}
\sigma &&
= - i {\frac{\lambda}{2}}\left \langle \left[ \psi^{\dag}, \tilde {\gamma}
^{0} \psi
\right]\right \rangle   \nonumber \\
&& ={ \lambda \over 2 \tau }    \int {dk_\eta \over 2 \pi} \left[ (1-2
N_+(k))
\phi_{k}^{+\dag} i \gamma^0
\phi_{k}^+  + (2 N_{-}(k) -1)  \phi_{k}^{- \dag} i \gamma^0  \phi_{k}^-
\right] ~.
\label{gap2}
\end{eqnarray}
Later we will choose the initial state number densities $N_\pm (k)$ to be
Fermi-Dirac distributions  at a given $\mu$ and $T$. We remark here
that the $k$ in $N_\pm (k)$ refers to $k_\eta$ canonically conjugate
to $\eta$.  When we calculate the expectation value of the energy momentum
tensor to identify the energy and pressure in a comoving system we will
find that the $N_\pm (k_\eta)$ corresponds to the comoving number densities
pertinent to relativistic hydrodynamics.
After some algebra we find
\begin{eqnarray}
R_k(u) = \phi_{k}^{+\dag} i \gamma^0 \phi_{k}^+  &&= -\phi_{k}^{ - \dag} i
\gamma^0
\phi_{k}^- =
|\dot{f}_k^+ |^2 + (\ts^2-k_{\eta}^2) |f_k^+ |^2+ i \ts  \left (
\dot{f}_{k}^+ f^{+\ast }_k
-\dot{f}^{\ast+}_k f_{k}^+\right) \nonumber \\
&& = |f_k|^2 \left[ \tOm_k^2 - \tom_k^2 + 2 \ts (\ts+\tOm_k) +
({\dot{\ts}+\dot{\tOm}_k
\over 2 \tOm_k})^2  \right] ~.
\label{eq:ru}
\end{eqnarray}

Eq. (\ref{gap2}) reduces to the vacuum expression for the gap when
everything becomes
time independent so that
\begin{eqnarray}
&&\tOm_k -> \tom_k ~,\nonumber \\
&&
Z_k(u) \equiv | f_k^{\pm}|^2  \rightarrow \left[ (2 \tom_k) (\tom_k+
\ts)\right]^{-1} ~, \\ \nonumber
&&  \left[ \tOm_k^2 - \tom_k^2 + 2 \ts (\ts+\tOm_k) +
({\dot{\ts}+\dot{\tOm}_k
\over 2 \tOm_k})^2  \right] \rightarrow 2 \ts ( \ts + \tom_k) ~,
\end{eqnarray}
which leads to the vacuum gap equation:
\begin{equation}
\sigma_0
=   \lambda     \int {dk_\eta \over 2 \pi}
 ~~~{\sigma_0 \over   \tom_k}  = \lambda     \int {dk \over 2 \pi}
 ~~~{\sigma_0 \over \omega_k} ~.
\label{gap3}
\end{equation}
In the time evolving case we obtain instead:
\begin{equation}
\sigma  ={ \lambda \over  \tau }    \int {dk_{\eta} \over 2 \pi}  (1- N_+(k)
-
N_{-}(k)) R_k(u)  ~,
\end{equation}
which in dimensionless form can be written as
\begin{eqnarray}
\tilde{\sigma}
&& = \lambda    \int {dk_{\eta}  \over 2 \pi}(1- N_+(k) - N_{-}(k)) R_k(u)
~.
\label{eq:sigdim}
\end{eqnarray}
We can simplify eq. (\ref{eq:ru}) for $R(u)$ by using eq. (\ref{eq:uga}) to
obtain \begin{equation}
R_k(u) =  1 - 2 k_{\eta}^2 ~~|f^+_k (u)|^2  \label{eq:ru2} ~.
\end{equation}

\section{Renormalization}

In this section we will show that the renormalization of the charge
$\lambda$
in the vacuum sector is sufficient to render the equation for $\sigma$
finite.
First let us remind ourselves that the effective potential for the
Gross-Neveu
Model can be written as:
\begin{equation}
\tilde{V} = {V\over N} = {\sigma^2  \over 2 \lambda } -{i\over 2} Tr
\ln(\gamma \cdot
\partial -
\sigma)~,
\end{equation}
where here the trace is only over the Dirac spinor indices.
The renormalized coupling constant defined at arbitrary $\sigma$ is given by
\begin{equation}
{d^2 \tilde{V} \over d \sigma d\sigma}= {1 \over \lambda_R(\sigma)} = {1
\over \lambda}
-
\int {dp \over 2 \pi} {1 \over \sqrt{p^2+\sigma^2}} +{1 \over \pi} ~.
\end{equation}
It is useful to define the logarithmically divergent integral
\begin{equation}
\Sigma(M^2) = \int_{-\Lambda}^\Lambda{dp \over 2 \pi}  {1 \over
\sqrt{p^2+M^2}}~.
\end{equation}
In particular if we choose the renormalization point to be at the minimum of
the potential, which also defines the fermion mass in this approximation,
one has
\begin{equation}
{dV \over d \sigma}|_{\sigma=m_f} =0 \rightarrow {1 \over \lambda} =
\int_{-\Lambda}^\Lambda  {dp \over 2 \pi} {1 \over \sqrt{p^2+m_f^2}} =
\Sigma(m_f^2) = {1 \over 2 \pi} \ln({\Lambda^2 \over m_f^2}) ~.
 \end{equation}
{}From the above equations, one deduces that
\begin{equation}
\lambda_R(m_f) = \pi ~,
\end{equation}
and also that
\begin{equation}
 { 1 \over \lambda_R(\sigma)} -{ 1 \over \lambda_R(m_f)}= \Sigma(m_f^2) -
 \Sigma(\sigma^2) = {1 \over 2 \pi} \ln({\sigma^2 \over m_f^2}) ~.
 \end{equation}

As derived earlier, for the evolution problem $\ts$ is given by:
\begin{eqnarray}
\tilde{\sigma}
&& = \lambda    \int {dk_{\eta}  \over 2 \pi}  (1- N_+(k) - N_{-}(k)) R_k(u)
\label{eq:sig}
 \end{eqnarray}

The apparent logarithmically divergent part of the momentum integral which
we call $\sigma^{div}$
comes from the term:
\begin{equation}
\ts^{div} =  \lambda     \int {dk_\eta \over 2 \pi} [1 - 2 \tom_k^2 Z_k(u)]
~.
\end{equation}

The logarithmic divergence can be isolated by doing
an adiabatic expansion of the integrand in the expression for $\ts^{div}$.
The first order adiabatic expansion of the equation for the
generalized mode functions $\tOm_k$ is obtained from the expression
for
$\tOm_k^2$ in  eq. (\ref{boost_WKB_D}) by replacing $\tOm_k,~ \dot{\tOm}_k$
by $\tom_k,
~\dot{\tom}_k$ in the left hand side of that expression.
We further use the expression for $\tom_k$ in terms of $\ts$
expressed in eq.(\ref{boost_omega_D}) to get:
\begin{equation}
\dot{\tom}_k
= \frac{\ts}{\tom_k} \dot{\ts} ~ ; ~
\ddot{\tom}_k  = \ddot{\ts} \frac{\ts}{\tom_k} +
\frac{\dot{\ts}^2}{\tom_k} \left[ 1 -
\left(\frac{\ts}{\tom_k}\right)^2 \right] ~.
\end{equation}
At large momentum, we therefore obtain the expansion:
\begin{equation}
\frac{\tOm_k^2}{\tom_k^2} =
1 - {1 \over 2 \tom_k^3}\left[ \ddot{\ts}
\left( 1 + \frac{\ts}{\tom_k} \right) +
\frac{\dot{\ts}^2}{\tom_k} \left( 1 -
\left(\frac{\ts}{\tom_k}\right)^2 \right) \right] +
{\dot{\ts}^2 \over 2 \tom_k^4} \left(1+\frac{\ts}{\tom_k}\right)
\left(1+\frac{3 \ts}{\tom_k}\right) \ldots ~.
\end{equation}
It therefore follows that
\begin{equation}
\ts^{div} =  \lambda   \int {dk_{\eta} \over 2 \pi} \left[\frac{\ts}{\tom_k}
\right] ~.
\end{equation}

We can renormalize the equation for $\ts$ be appropriately subtracting
this quantity when we renormalize the coupling constant, {\it or} we
can recognize that this divergence is only apparent once we utilize the
gap equation for the vacuum sector:

\begin{equation}
 \lambda^{-1} =  \int {dk_{\eta} \over 2 \pi}
 \frac{1}{\sqrt{k_\eta^2 + \tilde{m}_f^2}}= \int {dk \over 2 \pi}
\frac{1}{\sqrt{k^2+m_f^2} } \label{eq:lamr}
\end{equation}
That is we just to use the mode sum version of eq.(\ref{eq:lamr})
in place of $\lambda$ in eq. (\ref{eq:sig}) and keep enough
modes in the
numerator and denominator until the answer is independent of the
number of modes.  This approach has the advantage that it also allows one to
verify that
the coupling constant
is flowing according to the continuum renormalization group flow as we
increase the cutoff.

To explicitly renormalize the gap equation, we consider the
quantity:
\begin{equation}
\tilde{\sigma} [{1 \over \lambda_R(\sigma)} -{1 \over \lambda_R(m_f)} ] = {1
\over
2 \pi}  \tilde{\sigma}(u) \ln  ({\tilde{\sigma}^2 \over \tilde{m}_f^2} ) =
\int
{dk_{\eta}  \over 2 \pi}[ R_{f~k}(u) - ( N_+(k) + N_{-}(k)) R_k(u) ],
\label{eq:finitegap}
\end{equation}
where  $R_{fk}(u)$ is given by
\begin{equation}
R_{fk}(u) \equiv R_k(u) - {\ts  \over \tom_k}~.
\end{equation}
{}From the above discussion of the divergence structure, this equation
is manifestly finite.

\section{Initial Conditions}

To solve for the proper time evolution of this system one has a to solve
eq. (3.26)
\[
\left(- \frac{d^2}{d u^2}-
\tom_k^2  \pm  i~~{d \ts \over d u}  \right )
f^{\pm}_{k}(u) = 0 ~,
\]
as well as the eq. (\ref{eq:finitegap}) for the order parameter
\[
{1 \over
2 \pi}  \tilde{\sigma}(u) \ln  ({\tilde{\sigma}^2 \over \tilde{m}_f^2} ) =
\int
{dk_{\eta}  \over 2 \pi}[ R_{f~k}(u) - ( N_+(k) + N_{-}(k)) R_k(u) ] ~.
\]

To solve eq. (\ref{eq:finitegap}) requires knowledge of the initial
distribution of
fermions and
antifermions  $N_+(k)$ and $N_-(k)$.  In general the only conditions needed
on the
distributions
$N_+(k)$ and $N_-(k)$ are that they lead to  finite number and
energy densities.
The condition
on the mode functions for the energy density to remain finite at all times
is that
the high frequency modes must converge to their free field vacuum values.
Since the low
frequency modes do not effect the renormalization they can be chosen
arbitrarily.
Here we will choose the initial modes to match up with those found in the
WKB approximation initially. This will allow us to  introduce  an adiabatic
number
operator which has the added feature that it will interpolate from the
initial
distribution of $N_\pm(k)$ to the final outstate distribution functions.
Other choices
are perfectly acceptable but would lead to a jump in the number of
particles from their
initial values right after the initial time.
 Since we are
interested in exploring
the phase diagram on both sides of the tricritical regime, we
will choose our initial
state to be described by a value of $N_{\pm}(k)$   consistent with local
thermal and chemical
equilibrium described by the parameters
$T$ and $\mu$   (here the plus sign corresponds to the Fermion distribution
function and the minus sign to the antiFermion distribution).
\begin{equation}
N_{\pm}(k) = {1 \over e^{( \omega_k \mp \mu) /T}+1 } \label{eq:npm} ~,
\end{equation}
where
\[ \omega_k = E = \sqrt{k^2+\sigma^2(u=0)} = { 1 \over \tau_0}
 \sqrt{k_{\eta}^2 +\ts^2(u=0)} ~. \]
In LOLN approximation, the initial value problem is totally specified once
  $N_\pm$ and
$f_k$ and $\dot{f_k} $ are given.
The dynamics of the evolution is such that
the gap equation is obeyed at all times. If we start
in the regime where symmetry is broken, then  the initial value of the
Fermion mass obeys the equation \[
{\sigma }(u=0)  =  {\ts(u=0)  \over  \tau_0} ~,
\]
where $\ts $ satisfies the gap equation eq. (\ref{eq:finitegap}) with $u=0$.
If we start our simulation in the unbroken mode then we have instead
\[ \ts(u=0) =0 ~. \]
We notice that in LOLN when $\ts \equiv 0$ the Fermions evolve as if
they are free and
massless.   In the full theory, one obtains non-trivial dynamics for the
unbroken phase
through fluctuations that are ignored in the LOLN approximation.  Thus in
LOLN we need to
think of the $\sigma=0$ case as a limit of taking initial conditions with
small but
finite initial
mass  in order to have nontrivial dynamics.
As we will see below, finding the limit of zero mass is numerically well
defined once
initial masses  are
$\leq 10^{-6}~ m_f$. For our simulations an initial value of $\sigma(0) =
10^{-7} m_f$
is used.

It is convenient to choose the initial $ \tau_0 ={1 \over m_f}$ and measure
the
proper time in these units.
The adiabatic initial conditions on the mode functions $f$ which we will
discuss
in more detail later correspond to
\begin{equation}
\tOm_k(0) = \tom_k(0); ~~~ \dot{\tOm}_k(0) = \dot{\tom}_k(0)={\ts \over
\tom_k} \dot{\ts}~,
\end{equation}
so that using eq. ({\ref{eq:fdot}) we have
\begin{eqnarray}
f_k(0) && = {N_k \over \sqrt{2 \tom_k}} ~,\nonumber \\
\dot{f}_{k}^+ (0)&& = \left(- i  \tom_k- { \dot{\tom}_k + \dot{\ts} \over
2 \tom_k}
\right)  f_{k}^+ (0) ~, \nonumber \\
N_k^2 && = { 2 \tom_k(0) \over \left[{2 \tom_k^2(0) +2 \tom_k(0) \ts(0)}+
\left( {\dot{\ts}(0)+ \dot{\tom}_k(0) \over 2 \tom_k(0)}
\right)^2 \right]}~.  \label{eq:initial}
\end{eqnarray}

We now  show  that  $\dot{\ts}(u=0) =0$ is
required if we choose  adiabatic initial data for the mode functions.
For adiabatic initial conditions
\begin{equation}
\dot{\ts}(0) =  \lambda \int {dk_{\eta} \over 2 \pi}N_k^2 { k_{\eta}^2
\over  \tom_k(0)}
\left[ {\dot{\ts} + \dot{\tom}_k(0) \over \tom_k} \right] \biggl(1- N_+(k)
-N_{-}(k) \biggr) ~.
\label{eq:sigmadot}
 \end{equation}
Eq. (\ref{eq:finitegap}) and eq. (\ref{eq:sigmadot})  are two
equations for
the intial conditions $\ts$ and ${\dot \ts}$ in terms of $\mu$ and $T$.
However we realize that the integral for ${\dot \ts}(0)$ is proportional
to  ${\dot \ts}(0)$ times the bare coupling $\lambda$ times a finite
function of $\ts(0),{\dot \ts}(0), \mu, T$.  Since the bare
coupling goes to zero with the cutoff, the only value of ${\dot \ts}(0)$
consistent with the adiabatic initial conditions is
\begin{equation}
{\dot \ts}(0) = 0~.
\end{equation}
It then follows that
\begin{equation}
R_k(u=0) = {\ts(0) \over \tom_k(0)}~,
\end{equation}
so that the equation for $\ts(0)$ simplifies
to
\begin{equation}
{1 \over 2 \pi}  {\ts}(0) \ln  ({\tilde{m}_f^2 \over
\ts^2(0) }) = \int {dk_{\eta}  \over 2 \pi}\biggl( N_+(k) + N_{-}(k)\biggr)
R_k(0)
~.  \label{eq:finitegap2}
\end{equation}
This is equivalent to the gap equation arising from the  Effective
Potential at finite  chemical  potential and temperature
\begin{equation}
\ln {m_f^2 \over \sigma^2 }=  \int_{-\infty}
^{\infty} {dk \over \omega_k} [ 2- \tanh{\omega_k+
\mu \over 2 T} - \tanh{\omega_k -\mu \over 2 T} ]  ~.\label{eq:sigma3}
\end{equation}
if we choose $N_\pm(k)$ to be equilibrium Fermi-Dirac distributions.

Summarizing, for our choice of adiabatic initial conditions our mode
functions
initially are:
\begin{eqnarray}
f_k(0) && = {N_k \over \sqrt{2 \tom_k}};~~  N_k^2
 = [\tom_k(0) +\ts(0)]^{-1}~, \nonumber \\
\dot{f}_{k}^+ (0)&& = - i  \tom_k   f_{k}^+ (0)~. \nonumber \\
 \label{eq:initial2}
\end{eqnarray}

We are interested in studying evolutions with and without phase transitions,
as
well as comparing the effects of traversing first vs. second order phase
transitions,
including the special case of traversing the tricritical point.  We
have thus chosen four separate illustrative  starting points on the phase
diagram of Fig. 1
for our numerical simulations, all assuming initial local thermal and
chemical
equilibrium.
For the case of no phase transition, case (1), we have chosen the starting
point
 ($ \mu_0= .2; T_0 = .3$). For typical initial conditions
for which the second order
phase transition is traversed (case (2)), we choose  $ \mu_0= .5; T_0 = .5$.
For the initial conditions $ \mu_0= .6; T_0 = .32$ (case (3))  the
system traverses the tricritical point.
Finally, for a typical case where the system undergoes a first
order phase transition, case (4) we choose ($ \mu_0= .8; T_0 = .3$).

We will find that
before the phase transition, the system can be described in terms of a
number
distribution with a proper time evolving temperature and chemical potential.
However, after undergoing a phase transition, the adiabatic single particle
distribution functions for Fermions and antiFermions (defined below) are far
from equilibrium and cannot be described by a chemical potential and
temperature which are independent of the momentum.

In performing our numerical simulations, we place the system in a box of
dimensionless
length $\tilde{L} = m_f L$,
and choose antiperiodic boundary conditions for the Fermions. That is
we let
\begin{equation}
k_\eta \rightarrow k[n] ={2 \pi (n-1/2) \over \tilde{L}} ~,
\label{eq:discrete}
\end{equation}
with $ n= -N, \cdots N$.  In our simulations, for the system to be in a
regime where the coupling
constant flows according to the renormalization group, one needs $N = 5000$.
The number of modes
increases if we want to study the really long time behavior of this system.
However, at these long times we
expect hard processes which allow rethermalization, and which are neglected
in this mean field study to
become very important.

\section{Energy-Momentum Tensor}
To understand the hydrodynamical properties of the evolution of the plasma,
we need to evaluate
the  expectation value of the  energy momentum tensor in
the initial density matrix.   Because $\langle T_{\mu \nu} \rangle$
is diagonal in the $\eta,\tau$
coordinate system,  we can read off the  comoving pressure and energy
density from
its diagonal entries.   The energy-momentum tensor is defined via:
\begin{equation}
 T_{\mu \nu} = -{2 \over \sqrt{-g}}{\delta S \over \delta g^{\mu \nu}}~.
\end{equation}
Scaling out the factor of $N$, and using the equation of motion we have
\begin{equation}
{T_{\mu \nu} \over N} = {i \over 2}  \bar{\Psi} \tilde{\gamma} _{(\mu}
\nabla_{\nu)} \Psi -{i \over 2} \nabla_{(\mu} \bar{\Psi}
\tilde{\gamma}_{\nu)}
\Psi  - g_{\mu \nu} {\sigma^2 \over 2 \lambda}~.
\end{equation}
Here the parenthesis means keeping both terms in the symmetrization in
$\mu \nu$.
If we  rescale the fields $ \phi = \sqrt{\tau} \psi$ and
use the Fourier decomposition for the rescaled fields we find for the
expectation value of the unrenormalized  $T_{\tau \tau}$
\begin{eqnarray}
 {\tau^2~ \langle T_{\tau \tau} \rangle \over N} &&= {\tilde{\sigma}^2 \over
2 \lambda} - {i \over 4}
\int_{-{\tilde \Lambda}}^{{\tilde \Lambda}} {dk_{\eta} \over 2 \pi} [(2
N_+(k)-1)
(\phi^{+\dag}_k  \partial_u \phi^+_k - \partial_u \phi_{k}^{+\dag}
\phi_k^{+})
\nonumber \\
&& +(1-2N_{-}(k)) (\phi_{k}^{- \dag}
 \partial_u \phi_{k}^-  - \partial_u \phi_{k}^{-\dag} \phi^{-}_k)]~.
\end{eqnarray}
Expanding in terms of the mode functions $f^{\pm}_k$ we obtain:
\begin{eqnarray}
\epsilon(\tau) \tau^2  && \equiv
{\tau^2~ \langle T_{\tau \tau} \rangle \over N} \nonumber \\
&&  = {\tilde{\sigma}^2 \over 2 \lambda} -
\int_{0}^{{\tilde \Lambda}} {dk_{\eta} \over 2 \pi}
(1-N_+(k)-N_-(k)) [2 \tilde{\sigma} + 4 \tOm_k (\tom_k^2 - {\tilde
\sigma}^2) |f_k^+|^2 ] ~,
\end{eqnarray}
where
\[  \tOm_k |f_k^+|^2 \equiv {i \over 2} (f_k^+\partial_u f_k^{+\ast} -
f_k^{+\ast} \partial_u f_k^+ ) ~.\]
The energy density contains an infinite  (quadratically divergent)
cosmological
term:
\[  g_{00} K = -  \int_{0}^{{\Lambda}} {dk \over \pi}  k ~, \]
that needs to be subtracted by hand.  The remaining  logarithmic
divergence is eliminated by coupling constant renormalization.
The divergence structure can be analyzed using an adiabatic expansion of
$\tOm_k$ in
terms of $\tom_k$ and recognizing the high momentum behavior of $\tOm_k$
is given by
\begin{equation}
   \tOm_k = \tom_k [ 1+ 0 ({1\over \tom_k})^3] ~.
\end{equation}
so that
\begin{equation}
|f_k^+|^2 = {1 \over 2 \tom_k (\tom_k+ \tilde{\sigma})} [ 1+ 0 ({1\over
\tom_k})^4] ~.
\end{equation}

Thus the divergent terms in $ \langle T_{\tau \tau} \rangle $ are exactly
the terms that
appear
in the effective potential.  Namely
\begin{equation}
{\tau^2~ \langle T_{\tau \tau}^{div} \rangle \over N}  = {\tilde{\sigma}^2
\over 2 \lambda} - 2
\int_{0}^{{\tilde \Lambda}} {dk_{\eta} \over 2 \pi} \sqrt{k_{\eta}^2 +
\tilde{\sigma}^2 } ~.
\end{equation}
Making the rescalings:
\[  \tilde{\sigma} = \sigma \tau;~~ {\tilde \Lambda}= \Lambda \tau ; k_\eta
=
k \tau  ~, \]
one recovers the unrenormalized effective potential for the Gross-Neveu
model:
(See eq. (2.21) of \cite{ref:us}).
\begin{equation}
V = {\sigma^2 \over 2 \lambda} - 2 \int_0^{\Lambda} {dk \over 2 \pi} [
\sqrt{k^2 + \sigma^2} -k] ~,
\end{equation}
where here we have subtracted the (infinite) $\sigma$ independent
cosmological
constant term.
Next using the fact that the bare coupling and the cutoff and the physical
Fermion mass of the vacuum theory are related by  the gap equation:
\begin{equation}
{1 \over 2 \lambda} = \int_0^{\Lambda} {dk \over 2 \pi} {1 \over
\sqrt{k^2+m_f^2}}~,
\end{equation}
we obtain after subtracting the cosmological constant term
\begin{equation}
{\langle T_{ sub~ \tau \tau}^{div} \rangle \over N}= V =  \int_0^{\Lambda}
{dk \over 2 \pi}[ {\sigma^2
\over \sqrt{k^2+m_f^2}} - 2 \sqrt{k^2 + \sigma^2} + 2k] = {\sigma^2 \over 4
\pi}[ \ln {\sigma^2 \over m_f^2} - 1] + O({1 \over \Lambda^2})~.
\end{equation}
Here we have chosen the zeropoint of the Effective Potential  to
be
zero at the {\it maximum} $ \sigma=0$.  Thus in the true broken symmetry
vacuum
$\sigma=0$, we have
\begin{equation}
V[ \sigma=m_f] = -{m_f^2 \over 4 \pi}~.
\end{equation}
Thus we expect (and will find) that when the system goes through the phase
transition into the broken symmetry phase, that the energy density will
approach this {\it negative}  true vacuum value.

The manifestly finite expression for  $\epsilon(\tau)
\tau^2$ is \begin{eqnarray}
\epsilon(\tau) \tau^2 &&=
\int_{0}^{\tilde {\Lambda}} {dk_\eta \over 2 \pi} \left[ {\tilde{\sigma}^2
\over  \sqrt{k_{\eta}^2 + \tilde{m}_f^2}} + 2 (k_{\eta} - \tilde{\sigma}) +
4
\tOm_k (\tilde{\sigma}^2 - \tom_k^2) |f_k|^2 \right] \nonumber \\
&&+\int_{0}^{{\tilde \Lambda}} {dk_{\eta} \over 2 \pi} ( N_{+}(k)+N_{-}(k) )
\left[
2 \tilde{\sigma} + 4 \tOm_k( \tom_k^2 - \tilde{\sigma}^2) |f_k|^2
\right]~.\label{eq:eps}
\end{eqnarray}
We have for the
expectation value of the unrenormalized  $T_{\eta \eta}$
\begin{eqnarray}
{\langle  T_{\eta \eta } \rangle \over N}   &&= - {\tilde{\sigma}^2 \over 2
\lambda} +{i
\over
2} \int_{-{\tilde \Lambda}}^{{\tilde \Lambda}} {dk_{\eta} \over 2 \pi}~
(i k_{\eta}) ~ \left[  (2 N_+(k) -1)
 \phi_{k}^{+\dag} \gamma^0 \gamma^3 \phi_{k}^+   \right.
 \nonumber \\
&&\left.  +(1-2N_{-}(k)) \phi_{k}^{-\dag} \gamma^0 \gamma^3 \phi_{k}^-
\right] ~.
\end{eqnarray}
Expanding in terms of the mode functions $f^{\pm}$ we obtain:
\begin{equation}
 { T_{\eta \eta } \over N}  = - {\tilde{\sigma}^2 \over 2 \lambda} +
\int_{0}^{{\tilde \Lambda}} {dk_{\eta} \over 2 \pi}
(1-N_{+}(k)-N_{-}(k)) ~4~(\tilde{\sigma}+ \tOm_k) ({\tilde
\sigma}^2 -\tom_k^2 ) |f_k^+|^2 ] ~.
\end{equation}
Multiplying by $\tau^2$ and keeping the lowest order in the adiabatic
expansion
as before we obtain that the divergent part of the pressure is given by
\begin{equation}
p_0 = - \int_{0}^{\Lambda} {dk \over 2 \pi} {2 k^2  \over \sqrt{k^2 +
\sigma^2}} = - {\Lambda^2 \over 2 \pi} - {\sigma^2 \over 4 \pi} \left( \ln
{\sigma^2 \over 4 \Lambda^2}  + 1 \right)~.
\end{equation}

This is to be compared with divergent part of the energy density   given by:
\begin{equation}
\epsilon_0 =  - 2 \int_{0}^{\Lambda} {dk \over 2 \pi}  \sqrt{k^2 +
\sigma^2} = - {\Lambda^2 \over 2 \pi} + {\sigma^2 \over 4 \pi} \left( \ln
{\sigma^2 \over 4 \Lambda^2} - 1 \right)~.
\end{equation}

As we discussed in \cite{covariant}, the momentum cutoff $\Lambda$ acts as a
noncovariant point splitting regulator, giving terms in the regulated
$\langle T_{\mu \nu} \rangle$ proportional to $\delta^i_{\mu}
\delta^j_{\nu}$
in which the spatial directions $i,j = 1,2,3$ are distinguished.  Since
these
terms do not appear in the $ \mu = \nu = 0$ time component, the energy
density
has the correct $\Lambda$ dependence and requires no correction to make it
agree with covariance.  However, because covariance requires that the
cosmological term is proportion to $g_{\mu \nu}$, the correct regulated
pressure must satisfy
\begin{equation}
p_0^\prime \equiv - \epsilon_0 ~.
\end{equation}
We need to enforce this condition {\it by hand} by adding the difference
\[ p_0^\prime - p_0 = - \epsilon_0 - p_0 \]
to $p_0$ which corrects for the noncovariant term induced by our
momentum
cutoff.  We then need to subtract off the correct cosmological term to
eliminate the quadratic divergence. As in the case of the
energy density, the  apparent logarithmic
divergence gets cancelled by  coupling constant renormalization.
We then find:
\begin{equation}
p_0^{\prime} - {\sigma^2 \over 2 \lambda} = -{\sigma^2 \over 4
\pi}[ \ln {\sigma^2 \over m_f^2} - 1]+ {\Lambda^2 \over 2 \pi}~.
\end{equation}
Thus the renormalized and ``covariantized" expression for the pressure is:
\begin{eqnarray}
p^{\prime}  \tau^2 &&= \int_{0}^{{\tilde \Lambda}} {dk_{\eta} \over 2 \pi}
\left[ (1-N_{+}(k)-N_{-}(k)) ~4~(\tilde{\sigma}+ \tOm_k) ({\tilde
\sigma}^2 -\tom_k^2 ) |f_k|^2  \right . \nonumber \\
&& \left. + 2 {k_{\eta}^2 \over \sqrt{k_\eta^2 + \tilde{\sigma}^2}} + 2
\sqrt{k_\eta^2 +  \tilde{\sigma}^2} - 2 k_{\eta} -{\ts^2 \over
\sqrt{k_{\eta}^2 + \tilde{m}_f^2}}\right] ~. \label{eq:p}
\end{eqnarray}

\subsection {Local Equilibrium Hydrodynamical Picture}

In this subsection we will develop a  simple
local equilibrium hydrodynamical model with an ultrarelativistic equation of
state which we can compare with our field theory simulation.
This type of model was first put forth by
 Landau  \cite{Landau}.  In the hydrodynamical model, Landau assumed that
that the flow of energy and momentum following an ultra-relativistic high
energy collision of protons or heavy-ions  behaved as an ideal fluid flow
into
the vacuum with initial conditions connected to a highly Lorentz contracted
disc of
matter. A related idea due to Bjorken \cite{Bjorken} was to assume that
because of the
flatness of rapidity distributions in heavy-ion collisions, the hydrodynamic
flow should
posess invariance under boosts.  In either case one has approximately that
the fluid velocity
$v_z=z/t$ and all variables only depend on the fluid proper time $\tau$ and
are independent
of the fluid rapidity $\eta$. In this
hydrodynamic approach the dynamics are incorporated into the local
equilbirium
equation of state  $p = p(\epsilon)$.  For a one-dimensional relativistic
hydrodynamic flow the energy momentum tensor is assumed to be that of an
ideal fluid
\begin{equation}
T^{\alpha \beta} = p g ^{\alpha \beta} + (\epsilon+ p ) u^\alpha u^\beta ~.
\end{equation}
Here $\alpha,\beta$ correspond to Minkowski coordinates and
\[ u^\alpha = { v^\alpha  \over \sqrt{1-v^2}} ~. \]
The covariant conservation law of energy and momentum is
\begin{equation}
   T^{\alpha \beta} _{;\beta} = 0~.
\end{equation}
Introducing the thermodynamic relations:
\[
    \epsilon + p = Ts; ~~   d \epsilon = T ds ~,
\]
and assuming the scaling law
\[     v= z/t  ~,\]
one finds \cite{CFS} for an  equation of state  of the form
$p = c_0^2 \epsilon$  that energy momentum conservation leads
to the two equations:
\begin{eqnarray}
 &&\partial_{\eta} \ln T =0 ~,\nonumber \\
 && \partial_u \ln T + c_0^2 = 0~.
\end{eqnarray}
Thus for an ideal one-dimension fluid in the scaling regime  one
obtains:
\begin{eqnarray}
{T \over T_0} &&= ({\tau_0 \over \tau}) ^{c_0^2} ~, \nonumber \\
{\epsilon \over \epsilon_0} &&= ({\tau_0 \over \tau}) ^{1+c_0^2} ~.
\end{eqnarray}
The ultrarelativistic limit has $c_0^2 = 1$ (the speed of light in our units
where $c=1$)
for a one dimensional fluid. For that case
\begin{equation}
 {\epsilon \over \epsilon_0} = ({\tau_0 \over \tau}) ^{2};~~ {T \over T_0} =
({\tau_0 \over \tau})~.
\end{equation}
We will find from our numerical simulations, that if we start in the
massless
(unbroken symmetry) regime then indeed $p= \epsilon$ and this fall off
pertains
until one goes through the phase transition, after which the system is no
longer in local equilibrium.

When there is also chemical equilibrium, it can be shown \cite{ref:Landau2}
that for a relativistic fluid
\begin{equation}
{dT \over T} = {d \mu \over \mu} ~,
\end{equation}
so when that occurs we also have
\begin{equation}
{\mu \over \mu_0} =
({\tau_0 \over \tau})~.
\end{equation}

\subsection{Local Equilibrium equation of state and Single Particle
Distribution
Functions}
We can use the equations for the renormalized energy density
eq. (\ref{eq:eps}) and
pressure eq. (\ref{eq:p}) to determine the
equilibrium equations for
$p(\mu,T)$ and  $\epsilon(\mu,T)$ and thus the equation of state $p=
p(\epsilon)$ when one is in chemical and thermal equilibrium . In
equilibrium, $\sigma(\mu,T)$ is
 zero if we are above the phase transition in the unbroken symmetry
phase. If we are in the broken phase, it is instead  given by
the solution of the renormalized gap equation eq. (\ref{eq:sigma3}).  The
mode
functions are given by:
\begin{equation}
   |f_k|^2 \tau^2 = { 1 \over 2 \omega_k (\omega_k+ \sigma)}~.
\end{equation}
In the unbroken symmetry regime, where $\sigma=0$, this simplifies to
\[  |f_k|^2 \tau^2 = { 1 \over 2 k^3}  ~.\]
For the unbroken symmetry case,the equations for $p$ and $\epsilon$ become
the same  ($p=\epsilon$)
and are given by:
\begin{equation}
p=\epsilon= \int {dk \over 2 \pi} ~~2 k~~ [N_+(k,\mu,T) + N_{-}(k,\mu,T)] ~,
\label{eq:equileos}
 \end{equation}
where $N_+$ and $N_-$ are given by eq. (\ref{eq:npm}) and we have set
$\sigma=0$.
In order to get a quantitative estimate of what will happen in our numerical
simulations, we will
assume that the system stays in local equilibrium as it evolves with $T$
evolving as predicted
by the hydrodynamical model so that
$ T/T_0 = \tau_0 / \tau$.  We also find from our numerical simulations
that when the
initial chemical potential is below  the tricritical value the chemical
potential falls similar to the case of chemical equilibrium: i.e. $
\mu /\mu_0 =
\tau_0 /
\tau$ .
 Let us now show that when we are in local chemical and thermal equilibrium
$N_\pm(k_{\eta}, \tau)$ becomses independent of $\tau$.
Because  $\mu$ and $ T$ scale as ${1 \over \tau}$, and $ k ={k_{\eta} \over
\tau}$,  we find that the distributions for  $N_{\pm}$ is
\begin{equation}
N_{\pm}(k_\eta) = {1 \over  1+ \exp( {k_{\eta} \mp  \mu_0  \over T_0} )} ~,
\end{equation}
so if we  plot these distributions against $k_{\eta}$ they are
 independent of $\tau$ and
just depend on the initial values $\mu_0$ and $T_0$.
A plot of $N_{\pm}$  vs $k_\eta$ for the initial conditions of case (3),  $
\mu_0 =.6 $
and $T_0 = .33$, the initial
conditions for passing through the tricritical point is
shown in Figs.  \ref{fig:Nplusminus}.

\begin{figure}
   \centering
   \epsfig{figure=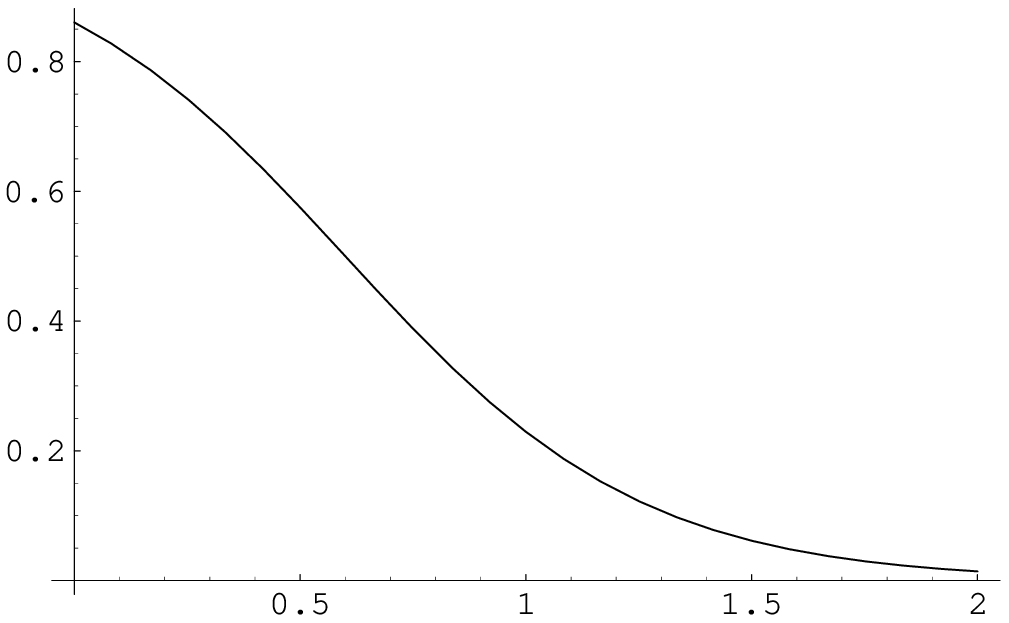,height=1.8in}
   \epsfig{figure=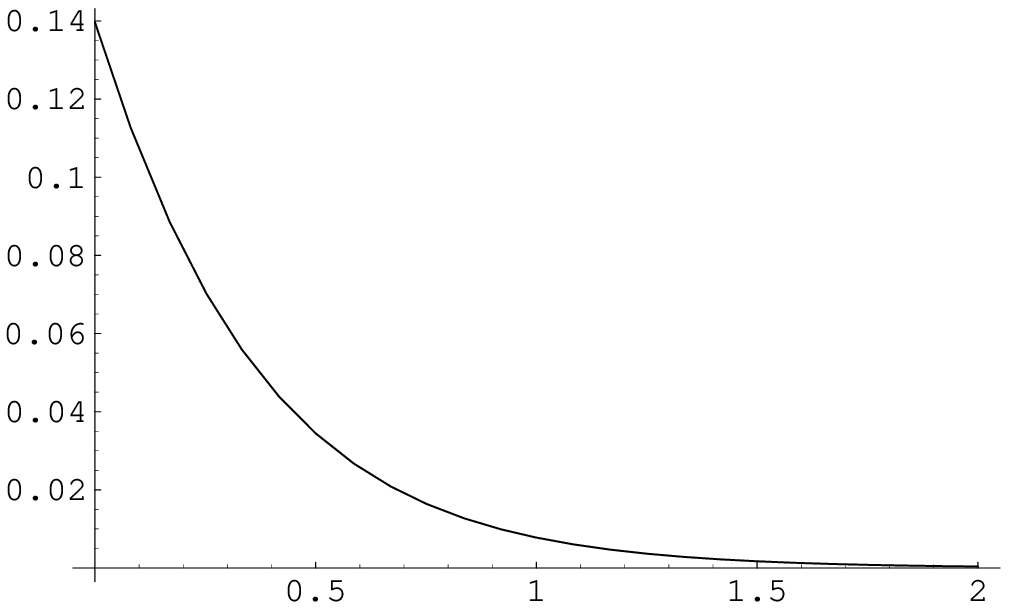,height=1.8in}
\caption{From left to right:
 $\tau$ Invariant thermal distribution $N_{+}$ vs. $k_\eta$  for initial
conditions
pertaining to traversing the tricritical point;
 $\tau$ Invariant thermal distribution $N_{-}$ for the same initial
conditions.
}    \label{fig:Nplusminus}
\end{figure}

{}From this discussion, we see that a deviation from the local equilibrium
hypothesis will show up
as a change in the adiabatic distribution functions for
$N_{\pm}(k_\eta,\tau)$
from its initial value as the
system evolves in
$u= \log{\tau}$. This deviation from will be greatest when the
trajectory undergoes a first order
phase transition. The behavior of the plasma ${\it
before}$ the phase transition is
very well described by hydrodynamics with equation of state $p=\epsilon$,
since
it is basically the evolution of a non-interacting relativistic gas
initially
assumed in equilibrium and described by:
\begin{equation}
p=\epsilon= {1 \over \tau^2} \int {dk_\eta \over 2 \pi} ~~2 k_\eta~~
[N_+(k_\eta) + N_{-}(k_\eta)] ~.
 \end{equation}
 This evolution in the massless phase is shown in fig. \ref{fig:etherm}.
which displays the expected ${1 \over \tau^2}$ falloff.

\begin{figure}
   \centering
   \epsfig{figure=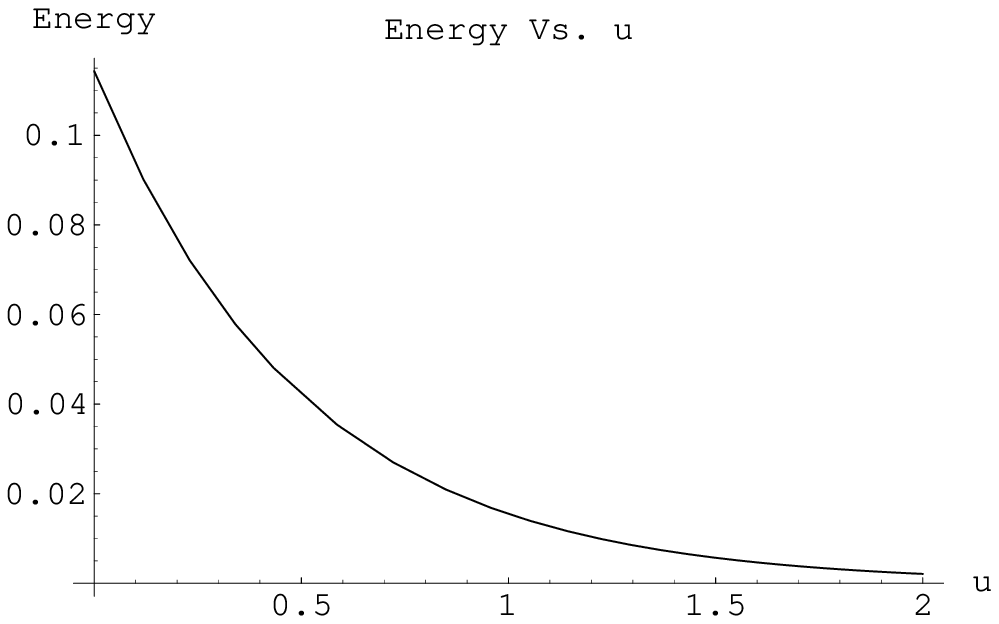,height=1.8in}
   \epsfig{figure=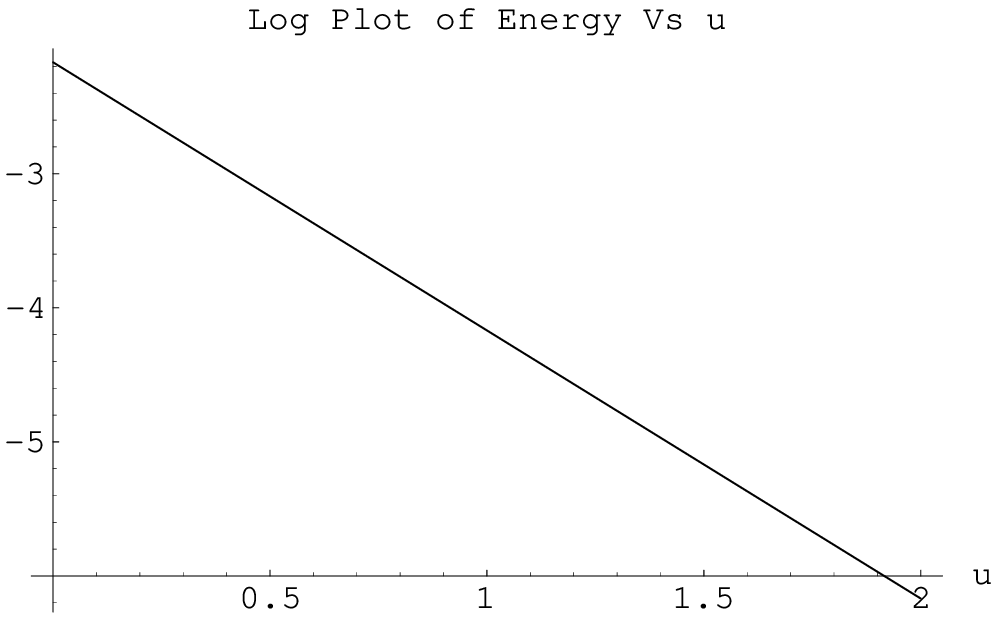,height=1.8in}
\caption{From left to right:
evolution of the energy density as a function of $u$  in the
massless phase;
same evolution for the log of the energy density as a function of $u$.
}    \label{fig:etherm}
\end{figure}

 \section{Adiabatic Number
Operator}
One important question for RHIC physics is how the quark and antiquark
distribution functions change in time since these time evolving distribution
functions enters into calculations of the particle production rates for
both pions and dileptons.  Of especial interest is how these distributions
 for Fermions and anti-Fermions gets modified as the plasma
traverses the chiral phase transition. Since particle number is
not conserved during the
evolution of this system,
one looks for a quantity that interpolates from the initially given
distribution to that which obtains once interactions have ceased. For this
problem because of the expansion into vacuum, interactions are automatically
diluted at late times and one eventually reaches the broken symmetry vacuum
state. LOLN is a mean field theory, which is related to a
 field theory for a fermion  with a proper time  varying mass which
at late times reduces to a free field theory. For
this reason, it is possible to
introduce various time dependent number operators which at late times
reduce to the exact out state number operator.  The approach  which we
follow
here, namely to define adiabatic number operators, is essentially the
same as that used when one has the problem of
quantum fields in time evolving curved background spaces  (see for example
ref.
 \cite{Birrell_Davies}). The basic idea is to use the
WKB approximation
to define a class of adiabatic vacuums upon which to define time evolving
number operators.  In the curved space problem, the curvature automatically
introduces a time evolving mass term for the quantum fields. In our problem
the time evolving mass term must be self-consistently determined.

Within the WKB approach,  there are several
versions
of the interpolating number operator  which differ by higher terms in the
WKB
expansion of the equation for the generalized frequencies  $\tOm$ which
enter
in a WKB parameterization for the exact  mode function $ f_k^{\pm}$. We will
describe
the zeroth and first order number operators below.  Once we determine
 the time evolving interpolating number operators, we fit them at each $u$
  to a Fermi-Dirac
distribution.  We will find that before the phase transition it is possible
to define a temperature and a chemical potential which are relatively
independent
of momentum. However, once the phase transition is traversed, the
interpolating number operator does not at all resemble an equilibrium
distribution function.  To define the interpolating number operator we use a
complete
set of wave functions $y_k^{\pm}(u)$  which are related to the solution of
the exact mode function equation
at some order in the WKB expansion. The WKB expansion is detemined
from the  second order differential
equation  (eq. (3.39)) for the generalized
frequencies $\tOm_k$.

In general, if we introduce a new basis $y_k^{\pm}(u)$ which are complete
and
orthonormal, but which
do not satisfy the Dirac equation, then the number operators themselves
become time dependent and the expansion of the quantum field becomes
\begin{equation}
\Phi (x) = \int {d k_{\eta} \over 2 \pi}[a({k},u)
y^{+}_k(u)
+c^{\dagger}(k,u) y^{-}_{k}(u)  ] e^{i k_{\eta} \eta} ~.
\label{wkb_fieldD}
\end{equation}
This expansion is an alternative to the expansion in terms of
 the initial time creation and
annihilation operators
\[
\Phi (x) = \int {d k_{\eta} \over 2 \pi}[b({k})
\phi^{+}_k(u)
+d^{\dagger}({{k}}) \phi^{-}_{{{k}}}(u)] e^{i k_{\eta} \eta}~.
\]
For this new expansion as well as the new creation and annihilation
operators
to obey
the canonical anticommutation relations, the mode functions must satisfy
 the orthonormality
conditions:
\begin{equation}
y^{a \dag}_k(u) y^{b}_k(u) = \delta^{ab} ~,
\end{equation}
for $a,b=\pm.$
The two sets of creation and annihilation operators are related by
a transformation which preserves the canonical structure, namely the
Bogoliubov
transformations:
\begin{eqnarray}
 a(k,u) &&= \alpha_k(u) b(k) + \beta_k^\ast d^\dag(k) ~, \nonumber \\
c^\dag(k,u) && = - \beta_k(u) b(k) + \alpha^\ast_k d^\dag(k) ~,
\end{eqnarray}
with the condition
\begin{equation}
|\alpha_k |^2 + |\beta_k|^2 = 1 ~.
\end{equation}
Inserting the Bogoliubov transformation into eq. (\ref{wkb_fieldD}) and
identifying terms we obtain
\begin{eqnarray}
  \phi_k^+(u) &&= \alpha_k(u) y_k^+ - \beta_k(u) y_{k}^-  ~,\nonumber \\
\phi_{k}^-(u) &&= \beta^\ast_k(u) y_k^+ + \alpha^\ast_k(u) y_{k}^- ~.
\end{eqnarray}
We can project out the Bogoliubov coefficients using the orthogonality
of the $y$ or the $\phi$, namely:
\begin{equation}
\alpha_k = y_k^{+\dag} \phi_k^+ ;~~~~~~ \beta_k^\ast = y_k^{+\dag} \phi_k^-
~.
\label{eq:proj}
\end{equation}
If we choose our initial conditions so that $y= \phi$, then initially
\begin{equation}
\alpha_k = 1; ~~~~ \beta_k =0.
\end{equation}
For that choice, the adiabatic particle number density will agree
initially with
the initial time number density.
 The interpolating number
operators for the Fermions and  anti-Fermions are defined  by
\begin{eqnarray}
N_+(k,u) &&= \langle  a^{\dag}(k,u)  a(k,u) \rangle ~,\nonumber \\
N_-(k,u) &&= \langle  c^{\dag}(k,u)  c(k,u) \rangle~,
\end{eqnarray}
where the expectation value is taken in the initial density matrix
parametrized by
$N_+$ and $N_{-}$ defined earlier.
In terms of  $\beta$ we find that
\begin{equation}
N^{\pm}(k,u) = N^{\pm}(k) + [1- N_+(k) - N_{-}(k) ]|\beta_k(u)|^2 ~,
\label{eq:number}
\end{equation}
so that the total number of particles minus antiparticles is conserved.
Since $\beta(u=0) = 0$ for adiabatic initial data, and at late times we
expect
$\sigma$ to be independent of $u$ so that $N(k,u)$ becomes the out
number operator at late $u$, the $N(k,u)$ interpolate between the initial
and
final values of the average phase space number density of particles.

We also see that if there are no particles present initially, then
$|\beta(k,u)|^2$ gives the particle spectrum, and its derivative is related
to the rate of pair production.  When particles are present then the
presence of these particle inhibits further production because of  Pauli
blocking.

\subsection{Zeroth Order Adiabatic Number operator}

The zeroth order in WKB wave functions are obtained from eq. (3.36)
by ignoring all derivatives in eq. (3.39).  This yields
\begin{eqnarray}
g_{k}^\pm (u) =
 \frac {1}{\sqrt{2\tom_{k}}} \exp\left \{ \mp \int_{0}^{u}
 i\tom_{k} (u')
du'\right \} ~ ,
\label{wkb_zero}
\end{eqnarray}
One easily verifies that introducing the basis functions:
\begin{equation}
y^+_k = u_k e^{-i \int~ \tom_k du};~~~ y^-_{k} = v_{k} e^{i \int~ \tom_k du}
~,
\end{equation}
\begin{eqnarray}
u_k &&= { -i \gamma^{\mu} k_{\mu} +\ts \over \sqrt{2 \tom_k (\tom_k+\ts)}}~~
\chi^+  ~,\nonumber \\
v_{-k} &&= { i \gamma^{\mu} k_{\mu} +\ts \over \sqrt{2 \tom_k
(\tom_k+\ts)}}~~ \chi^- ~,
\end{eqnarray}
that the spinors are orthonormal:
\[  u^\dag_k v_k=0;~~  u^\dag_k u_k = v^\dag_k v_k =1 ~,
\]
which guarantees that the orthonormality condition  eq. (7.2) is satisfied.
Using the relationship
\[    \beta^\ast = y_k^{+ \dag} \phi_k^-  ~,  \]
one finds that
\begin{equation}
|\beta_k|^2 = k_{\eta}^2 { (\tOm_k - \tom_k)^2 + \Delta_k^2 \over 2 \tom_k
(\tom_k+\ts)~~~ [\tOm_k^2 +
\tom_k^2 + 2 \tOm_k \ts + \Delta_k^2]}~,
\end{equation}
with
\[  |\alpha_k|^2 = 1 - |\beta_k|^2 ~,
 \]
and $\Delta_k$ given by eq. \ref{eq:Del}.

\subsection{first order WKB interpolating number operator}

If we keep terms up to and including all first derivatives
in eqs. (3.36) and (3.39), we obtain
the first order WKB wave functions
 $g_k^{\pm}$ given by:
\begin{eqnarray}
g_{k}^\pm (u) =
 \frac {1}{\sqrt{2\tom_{k}}} \exp\left \{ \int_{0}^{u}
\left ( \mp i\tom_{k} (u')
- \frac {{\dot{\ts}
}(u')}
{2\tom_{k} (u')}
\right )
du'\right \}  ~.
\label{wkb_ansatz_D}
\end{eqnarray}
The $g_k $ obeys the first order differential equation
\begin{equation}
\dot{g}_{k}^\pm (u)  = \mp i [\tom_k \mp i \Delta_0] g_{k}^\pm (u)~,
\end{equation}
where  $\Delta_0$ is
\begin{equation}
\Delta_{k0} = {\dot \ts + \dot \tom_k \over 2 \tom_k } ~.  \label{eq:Del0}
\end{equation}
We decompose  $\Phi$  as follows:
\[
\Phi (x) = \int {d k_{\eta} \over 2 \pi}[a({k},u)
y^{+}_k(u)
+c^{\dagger}({{k}},u) y^{-}_{{{k}}}(u)   ] e^{i k_{\eta} \eta} ~,
\]
where now the $y^{\pm}_{{ k}}$ are given by
\begin{equation}
y^{\pm}_{{ k}}(u) = A_k(u) \left[-\gamma^0 {d\over d u}
- i {\gamma^{3}} k_{\eta}
 + {\tilde\sigma}(u ) \right] g ^{\pm}_{k}(u) \chi^{\pm} ~.
\label{wkb_mode_eq_f}
\end{equation}
One can then verify that the $y^{\pm}_{{ k}}(u)$ are orthonormal providing
we choose
\begin{equation}
 A_k^2(u) =  {2 \tom_k  \over 2 \tom_k (\tom_k+\ts) + \Delta_{0k}^2(u)}
\exp \left[{\int_0^u du ~~{\dot{\ts} \over \tom_k}}\right] ~.
\end{equation}
At time $u=0$ we have $\Delta=0$ and
\begin{equation}
 A_k^2(u=0) =  {2 \tom_k  \over 2 \tom_k (\tom_k+\ts)} = N_k^2 ~,
\end{equation}
so that the exact and adiabatic wave functions match up.
Again using the relationship
\[    \beta^\ast = y_k^{+ \dag} \phi_k^-  ~,   \]
one finds that
\begin{equation}
|\beta_k|^2 = k_{\eta}^2 { (\tOm_k - \tom_k)^2 + (\Delta_k-\Delta_{k0})^2
\over
 [2 \tom_k (\tom_k+\ts) + \Delta_{0k}^2] [\tOm_k^2 +
\tom_k^2 + 2 \tOm_k \ts + \Delta_k^2]} ~,
\end{equation}
with
\[  |\alpha_k|^2 = 1 - |\beta_k|^2~.
 \]
and $\Delta_k$ given by eq. \ref{eq:Del}  and $\Delta_{0k}$ given by
eq.\ref{eq:Del0},
so that
\[ \Delta_k-\Delta_{k0} = {\dot \tOm \over 2 \tOm} - {\dot \tom \over 2
\tom} ~. \]

\section{Results of Numerical Simulations}

We have solved the simultaneous equations eq. (\ref{eq:mode}) and eq.
(\ref{eq:sigdim})  numerically by discretizing the Fourier modes in a box of
dimensionless  length  $\tilde{L}$ using antiperiodic boundary conditions
for
the Fermion modes. Our initial conditions were described
earlier and are based on adiabatic initial conditions and an equilibrium
value
for $N_{\pm}$. We have varied the time step, length of the box, as well
as the number of modes until the answer was insensitive to our choices.
The sensitivity to some of these parameters will be displayed below.
 Since the phase
transition occurs near $u=2$ or $\tau=7.4$, we typically continue our
calculation until $u=4$ or $\tau= 54$.  In that regime of $u$,
it was sufficient
to choose $\tilde{L}=500$  and keep 5000 modes.  The time step needed for
these values was $ du = 0.00004$.
We use a fixed grid in the dimensionless momentum $k_\eta = k \tau$ of 5000
points.   This was sufficient to insure that the range of integration
in the calculation of $\sigma$ included physical momentum $k$ at least of
the
order $10 m_f$.  Because of our fixed grid in $k_\eta = k \tau$,
 we had to increase the number of modes we include in
our evaluation of $\sigma$ as the proper time increases.

In what follows we discuss some of the finite size  effects of our grids.
Because
of the exponential dependence of $\tau$ on $u$ the number of modes needed
for
an accurate answer at late times grows rapidly.  We can see the dependence
of
our answers at late proper times on the parameters $\tilde{L}$, the number
of
modes $N$ as well as the time step $\Delta u$ in the figures below.  If we
keep the number of modes fixed at 5000, and decrease $\tilde{L}$ then we
increase
the
momentum range in our integrals and improve the result for $\sigma$ at late
times.   This is seen in fig. \ref{fig:length}.
\begin{figure}
   \centering
   \epsfig{figure=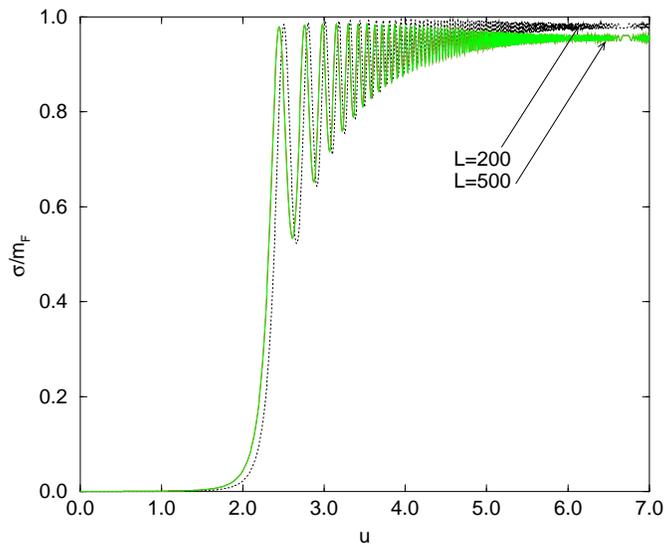,width=4.0in}
\caption{ Evolution  of $\sigma$ as a function of $u$ for
two different values of the dimensionless box length.}    \label{fig:length}
\end{figure}

In fig. \ref{fig:time} we show how the evolution of $\sigma$ depends on the
time step $du$.  We see that once we have a time step $du = 0.00004$
then our results are insensitive to any further reduction in time step.
\begin{figure}
   \centering
   \epsfig{figure=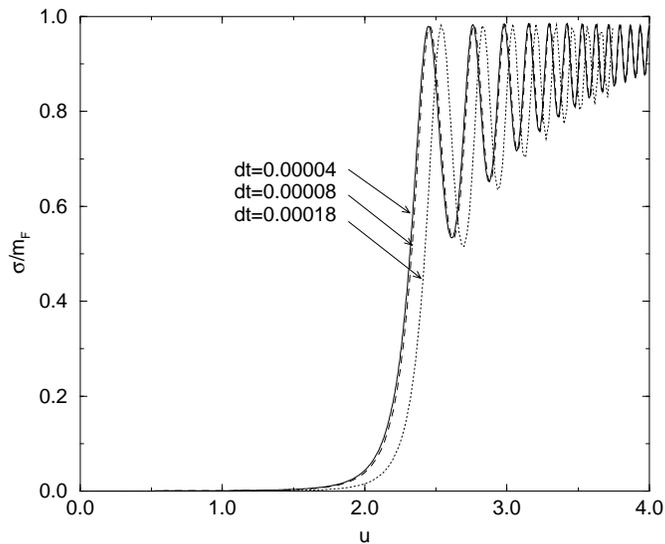,width=4.0in}
\caption{ Evolution  of $\sigma$ as a function of $u$ for
different values of $du$.}    \label{fig:time}.
\end{figure}

In fig. \ref{fig:points} we see how the approach to  the continuum value of
${\sigma \over m_f}=1$ depends on the number of Fourier modes.  We see that
at
late times there is still some depedence of the asymptotic value on the
number
of modes.

  \begin{figure}
   \centering
   \epsfig{figure=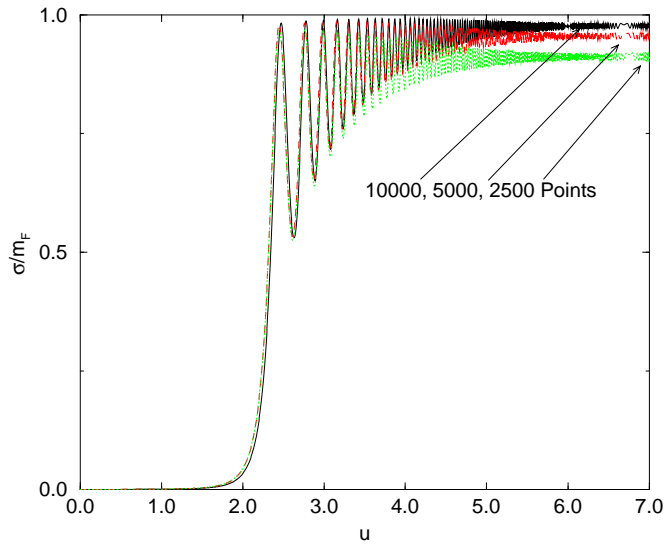,width=4.0in}
\caption{ Evolution  of $\sigma$ as a function of $u$ for
increasing numbe of Fourier modes.}    \label{fig:points}.
\end{figure}

Another issue we would like to address here is the dependence of the
evolution on the initial conditions chosen. We have seen that in LOLN,
if we start with the fermion mass {\it exactly} zero, then the theory is
non-interacting.  Thus we must consider the massless theory as the limit
of the massive theory, with the understanding that in higher order the
fluctuations ignored in LOLN will make the equation for the fermion modes
nontrivial even in the unbroken symmetry phase.   To see that the theory
actually approaches a limiting behavior, we considered initial condition
where either the mass is small and not zero or the time derivative of the
mass
is non zero.  The limiting theory is seen to be more readily accessed by
choosing $\dot{\sigma} =0$ and gradually letting $\sigma \rightarrow 0.$
For ${\sigma_0 \over m_f} \leq 10^{-7}$ the transition point changes
little.
We illustrate this graphically below.
Figure \ref{fig:sigdot} displays the time evolution for different
initial values of $\dot{\sigma}$.
Figure \ref{fig:sig} displays the time evolution for different
initial values of $\sigma$.
As we go to bigger initial  values of  $\sigma$ or $\dot{\sigma}$ the
transition is more a gradual crossover rather than a sharp transition.

\begin{figure}
   \centering
  \epsfig{figure=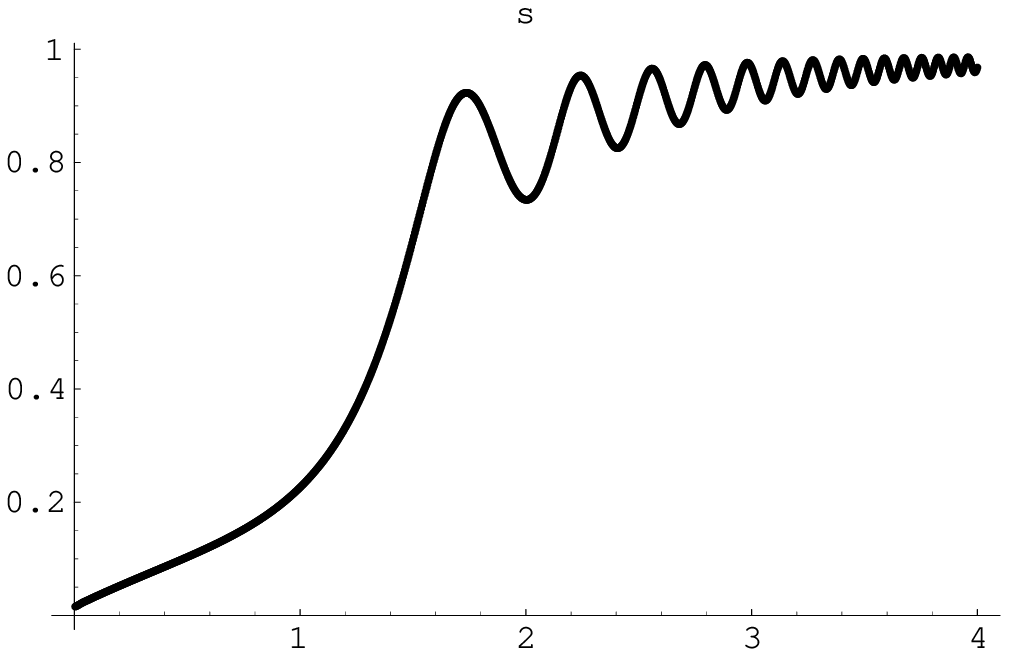,width=2.5in,height=1.8in}
   \epsfig{figure=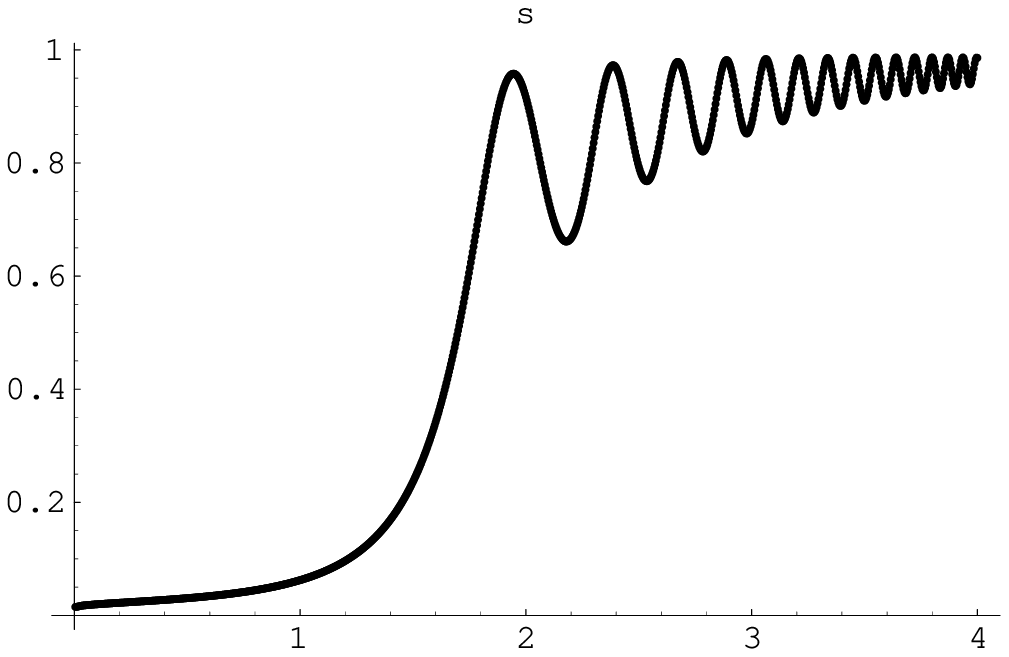,width=2.5in,height=1.8in}
   \epsfig{figure=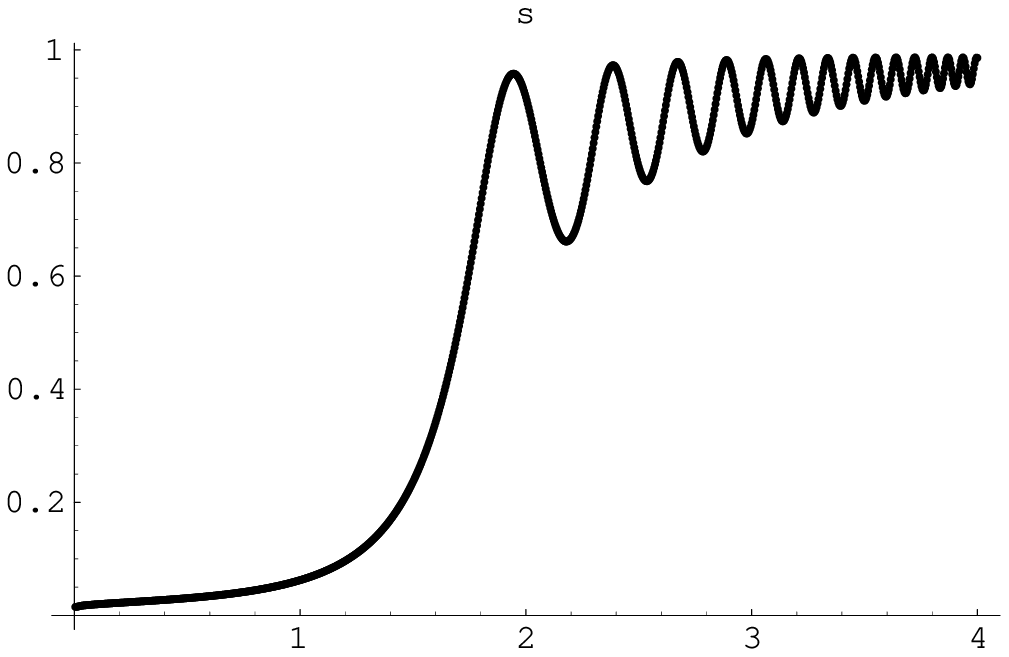,width=2.5in,height=1.8in}
\caption{ Evolution $\sigma$ as a function of $u$ for 3 different values of
$\dot{\sigma}$.
First figure is for $\dot{\sigma}=10^{-1}$ ,
second figure is for $\dot{\sigma}=10^{-4}$ and
third figure is for $\dot{\sigma}=10^{-5}$.
}    \label{fig:sigdot}
\end{figure}

\begin{figure}
   \centering
  \epsfig{figure=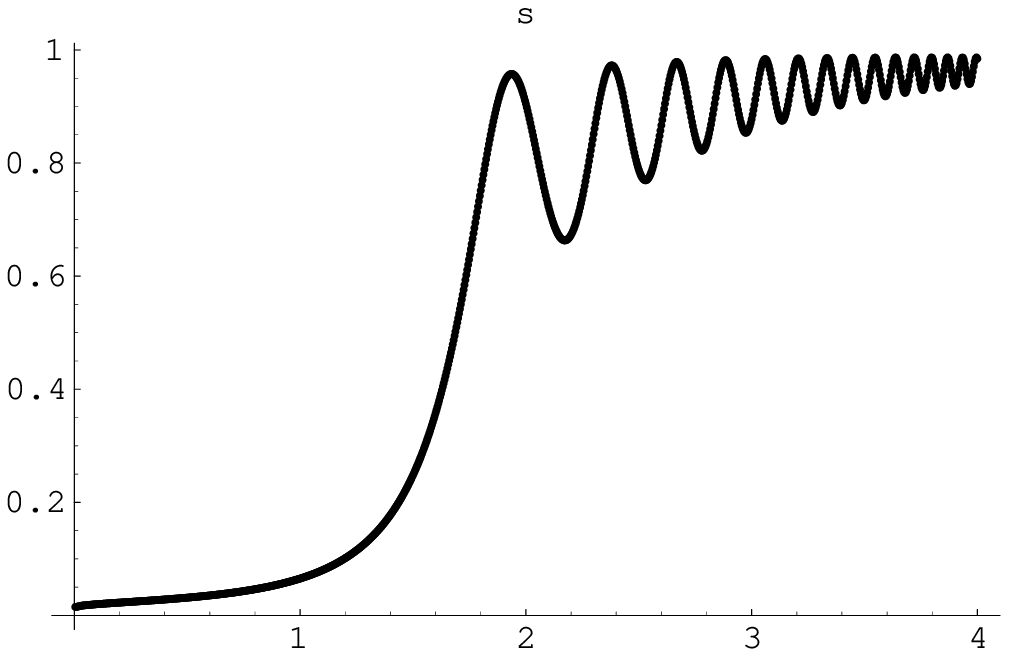,width=2.5in,height=1.8in}
   \epsfig{figure=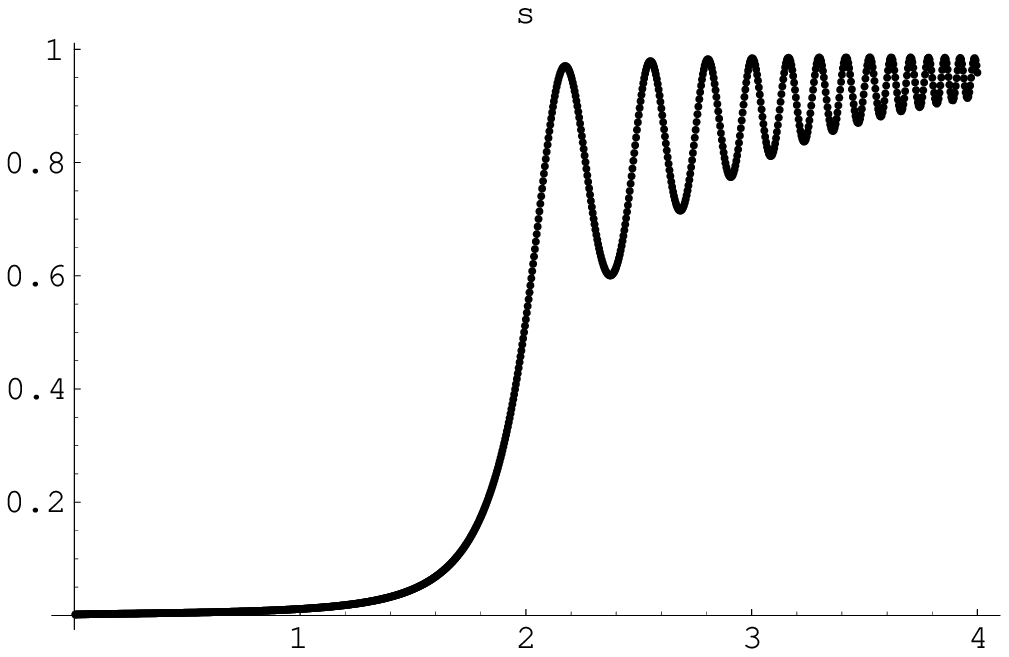,width=2.5in,height=1.8in}
   \epsfig{figure=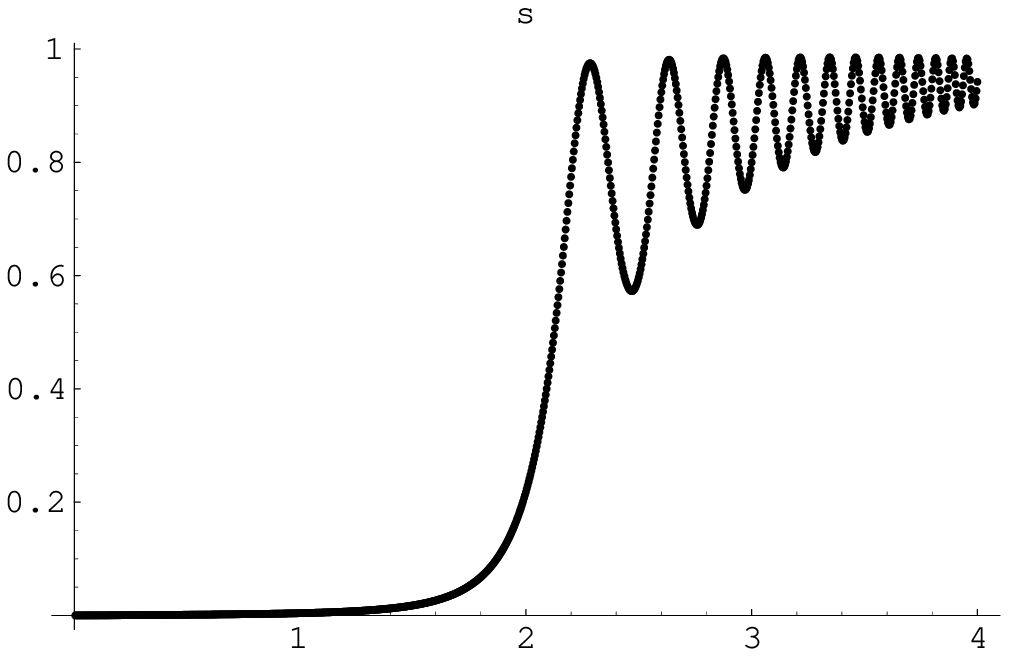,width=2.5in,height=1.8in}
\caption{ Evolution $\sigma$ as a function of $u$ for 3 different values of
$\sigma$.
First figure is for $\sigma=10^{-2}$ ,
second figure is for $\sigma=10^{-3}$ and
third figure is for $\sigma=10^{-7}$.
}   \label{fig:sig}
\end{figure}

Based on this discussion,
in the simulations presented below, we will canonically use the values
$ du = 0.00004$, $\tilde{L}=500$, $N=5000$, and $\sigma=10^{-7}$.

\subsection{Proper time evolution of $\sigma$,
$N_\pm$,$T$ and $\mu$}}

\begin{figure}
   \centering
   \epsfig{figure=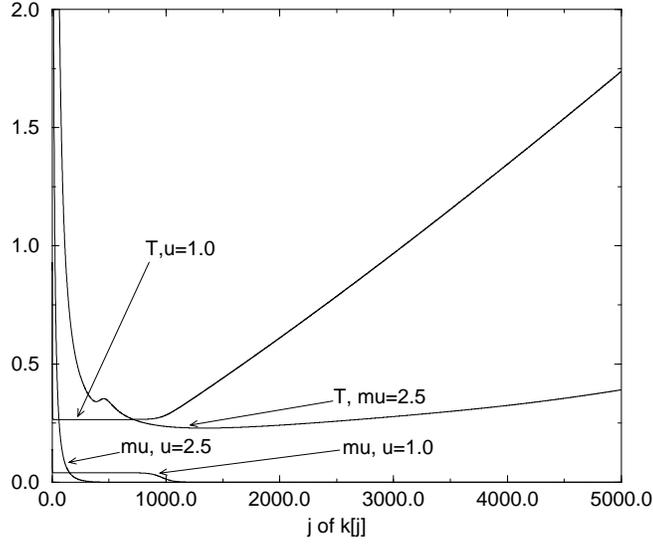,width=4.0in}
\caption{Effective temperature and chemical potential as a function of
$k_{\eta}$ for u=1.0 and u=2.5}    \label{fig:tmuk}
\end{figure}

We have determined
 $\ts(u)$ in terms of the mode functions using two different methods: the
 explicitly renormalized equation,  eq. (\ref{eq:finitegap}) and the
combined
set of eq. (\ref{eq:sig}) and eq. (\ref{eq:lamr}). The latter set of
equations
also allows one to check whether one has included enough mode functions for
the
coupling constant to flow logarithmically as in the continuum limit.
In the previous section we have discussed how these numerical results
depended
on various discretization parameters as well as the small initial value
of the the explicit symmetry breaking.   From the mode functions, using
eq. (\ref{eq:proj}), one determines the Bogoliubov coefficints and
then determines $N_\pm(k,u)$ from eq. (\ref{eq:number}).
By
comparing eq. (\ref{eq:number}) with an equilibrium paramatrization:
\begin{equation}
N_{\pm}(k,u) \equiv \left(1+ \exp[{\omega_k(u) \mp \mu(k,u) \over
T(k,u)}]\right) ^{-1}  ~, \end{equation}
where $ \omega_k = \sqrt{{k_\eta^2 \over \tau^2} + \sigma^2}$
we  then obtain two equations for the two parameters $T(k,u)$ and $\mu(k,u)$
 as a function of $k_\eta$. When these quantities are independent  of
 $k_{\eta}$ this
 defines
the proper time evolving temperature and chemical potential. Some
indication that an equilibrium parameterization is possible
in a 1+1 dimensional field theory evolution  in the
LOLN approximation was already shown in the work of Aarts et. al.
\cite{ref:Aarts}. In our simulation we find
that $T$ and $\mu$ are independent
of $k_\eta$ (except at high momentum) until the system undergoes the chiral
phase
transition.   A typical example of the dependence of $T$ and $\mu$ on
$k_\eta$ is shown in fig. \ref{fig:tmuk}.  For these initial conditions the
phase transition occurs
near  $u = 2$. The connection
between
$k_\eta$ and $j$ of the plot is given in eq. (\ref{eq:discrete}), namely:
$ k[j] = 2 \pi (j-1/2)/L. $

Because the system goes out of equilibrium after the phase transition,
determining
 $T$ and $\mu$ in that regime is somewhat arbitrary and we use a value
averaged
 over
$k_\eta$.
As we have shown earlier, if the system evolves in local
thermal equilbrium  in
a massless phase, then $T$ falls as $1/\tau$.  If there is also chemical
equilibrium then $\mu$ falls identically to $T$. We will
find this is precisely true before the phase transition when
there is a second order phase transition.  For the first order transition
$\mu$ appears to fall faster than $T$.   We also showed that if local
chemical and thermal equilbrium are maintained, then the
spectrum of
particles and antiparticles when plotted against the dimensionless momentum
$k_\eta= k\tau$ should
be independent of $\tau$. Thus any change in this spectra is an indication
of the system going
out of equilibrium.  We expect because of the latent heat released during a
first order transition that
the distortion of the spectra would be greatest in that case which is what
we will find below. For the time evolution of the single particle
distribution, we pick two time $u=1.5$ and $u=2.5$ which are before
and after the phase transition.

\begin{figure}
   \centering
   \epsfig{figure=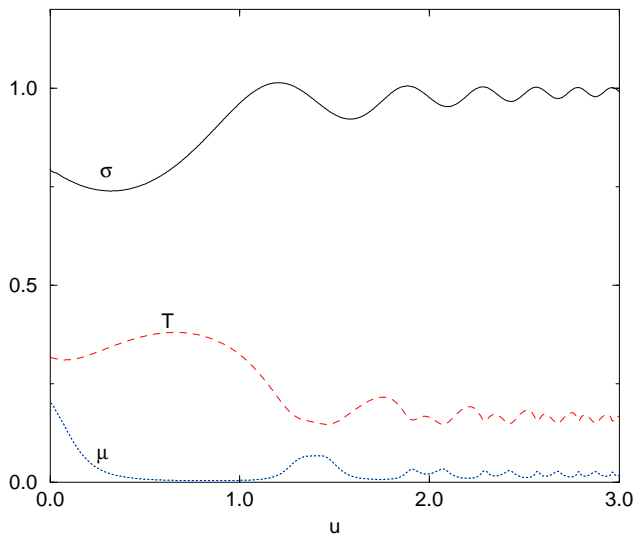,width=4.0in}
\caption{Typical evolution of $\sigma/m$, $\mu$ and $T$ as a function of $u$
for
initial conditions below the phase transition temperature}
   \label{fig:broken}
\end{figure}

 Let us now focus on
the four initial conditions described earlier and look simultaneously at the
time evolution
of
the order parameter $\sigma$, $T$ and $\mu$ as a function of
$u=\ln{m_f \tau}$. We separately plot the evolution of $N_\pm(k_\eta,u)$.
In all these plots all dimensionful parameters are
scaled by the mass of the Fermion $m_f$ in
the broken symmetry vacuum state. \newline
\noindent {\it case (1)} $\mu_0= .2; T_0= .3$ -- no phase transition\newline
Here we start below the phase transition so
that
the Fermions initially have non-zero mass. The results for the order
parameter $\sigma$ as well as for $T$ and $\mu$ are displayed
in Fig. \ref{fig:broken}.
We see here that although the chemical potential goes to zero as the plasma
expands, after $u=1.5$,  all the
parameters become relatively independent of $u$ and are
defined by an equilibrium freezeout temperature of  approximately $.2 m_f$.

\begin{figure}
   \centering
   \epsfig{figure=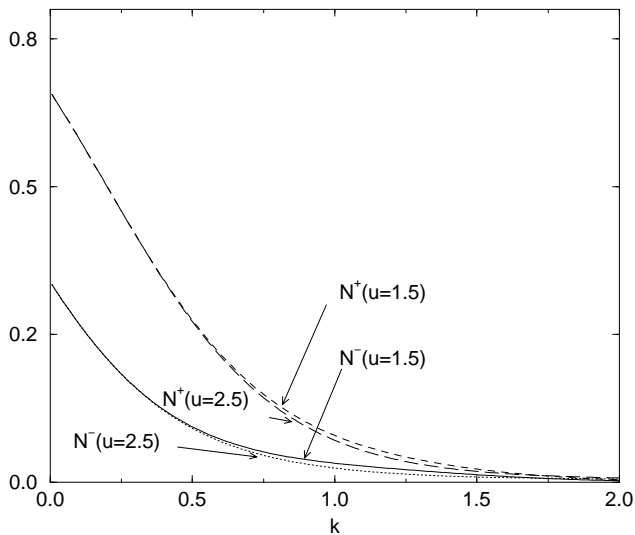,width=4.0in}
\caption{Evolution of $N_{\pm}$ as a function of $u$ when there
is no  phase transiton. The momentum displayed is $k_\eta= k \tau $.}
\label{fig:Nno} \end{figure}

For the broken symmetry case, the plasma
essentially stays in  equilibrium  throughout the evolution in that the
proper
time evolving  $N_\pm$
is relatively independent
of $\tau$ and maintains its Fermi-Dirac shape.  This
is seen in Fig. \ref{fig:Nno}.

\begin{figure}
   \centering
   \epsfig{figure=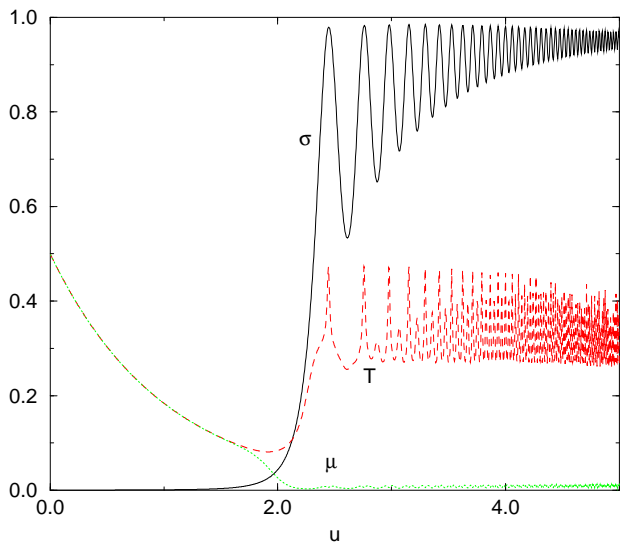,width=4.0in}
\caption{ Evolution of $T$,$\mu$ and $\sigma$ as a function of $u$ for
second-order
phase transition}    \label{fig:tmusig}
\end{figure}

\noindent {\it case (2)} $\mu_0= .5; T_0= .5$ -- second
order phase transition\newline
The results for starting in the ``unbroken" mode  and traversing a
second order phase transition are shown in  Fig. \ref{fig:tmusig}.

\begin{figure}
   \centering
   \epsfig{figure=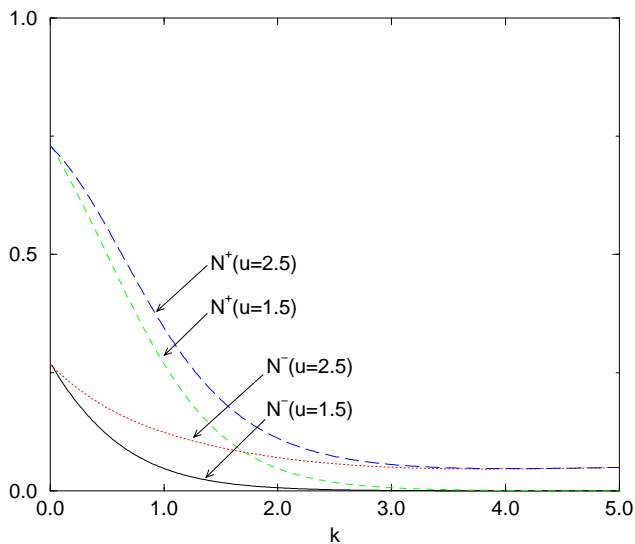,width=4.0in}
\caption{Evolution of $N_{\pm}$ as a function of $u$ when there
is a second order phase transiton. The momentum displayed is $k_\eta= k \tau
$}    \label{fig:Nsec} \end{figure}

We notice that $\sigma(\tau)$  shows a sharp transition
during evolution from the unbroken mode ($\sigma=0$) to the
broken symmetry mode. Before the phase transition the temperature
falls consistent with the equation of state  $p= \epsilon$. The chemical
potential follows
the temperature in that regime which means the system is also
in chemical equilibrium. After the phase transition, the chemical
potential goes to zero wheras the temperature freezes out at a $T \approx
.35m_f$.  After the phase transition, there is now a mass scale $m_f$ which
leads to oscillations of $\sigma$ around the final vacuum value. As
discussed
earlier, to obtain non-trivial dynamics we chose a small explicit symmetry
breaking parameter $\sigma/m_f = 10^{-7}$.

Going through a second order phase transition {\it does} produce
a noticable effect
in distorting the Fermi-Dirac distribution as shown in Fig. \ref{fig:Nsec}

\begin{figure}
   \centering
   \epsfig{figure=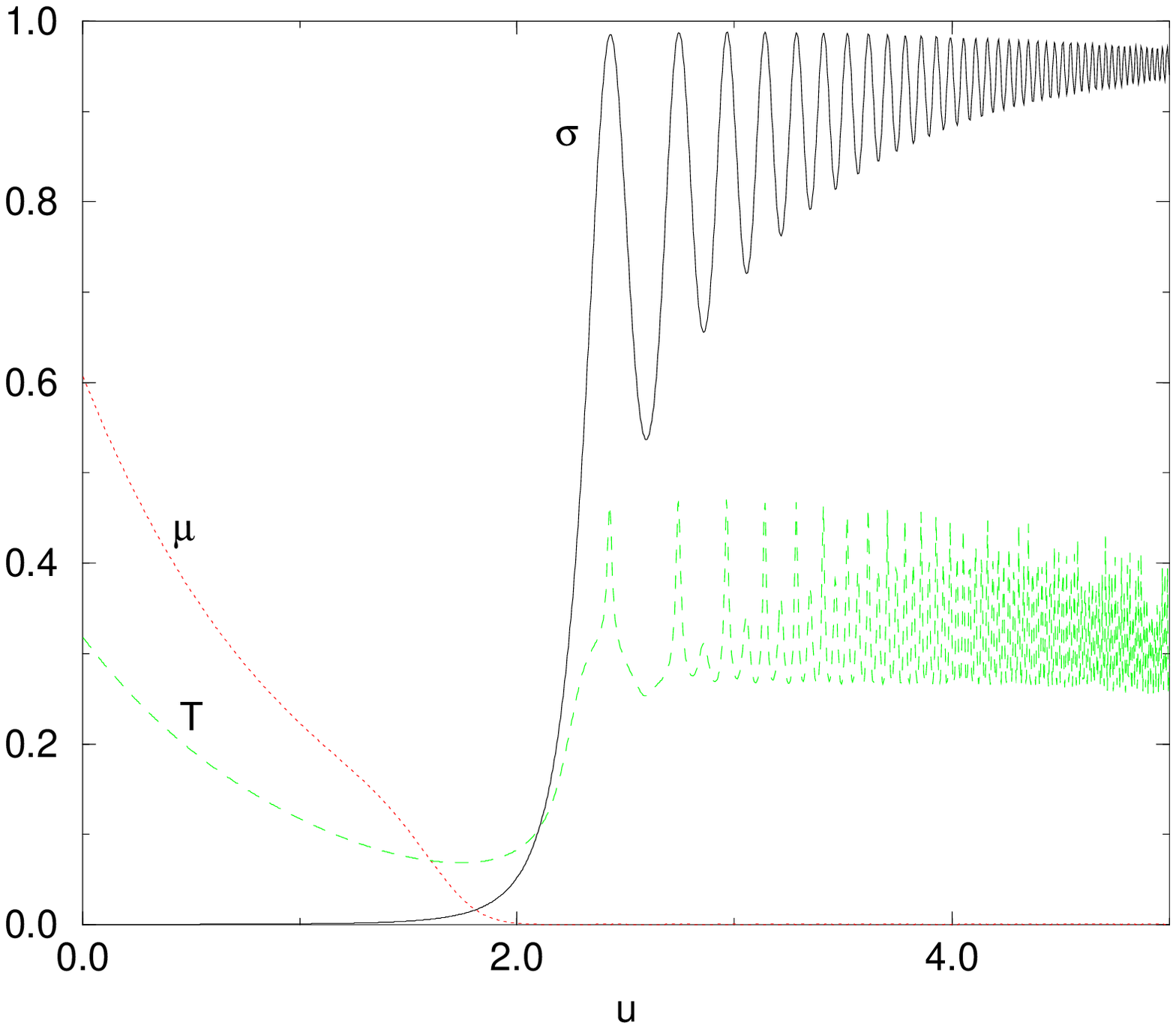,width=4.0in}
\caption{ Evolution of $T$,$\mu$ and $\sigma$ as a function of $u$ passing
throught the tricritical point }
   \label{fig:tricrit}
\end{figure}

\noindent {\it case (3)} $\mu_0= .6; T_0= .32$ --traversing the tricritical
point \newline
The results for this numerical simulation are shown in Fig.
\ref{fig:tricrit}.

\begin{figure}
   \centering
   \epsfig{figure=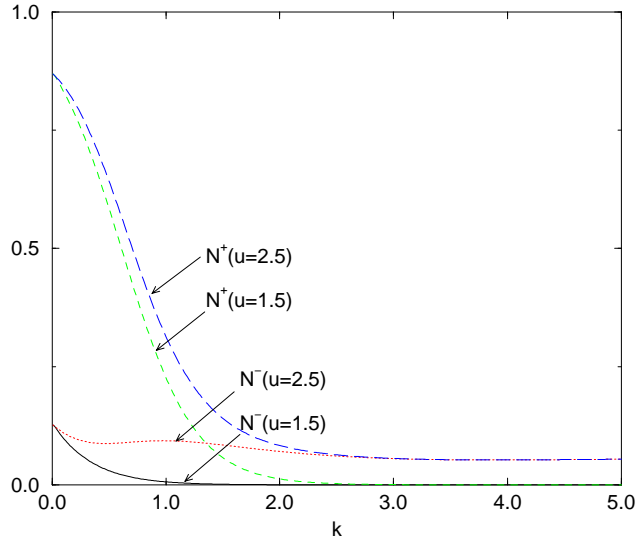,width=4.0in}
\caption{Evolution of $N_{\pm}$ as a function of $u$ when one
traverses the tricritical point.The momentum displayed is $k_\eta= k \tau$.
}
  \label{fig:ntri} \end{figure}

We notice that $\sigma(\tau)$  again shows a sharp transition
during evolution from the unbroken mode ($\sigma=0$) to the
broken symmetry mode and again the temperature
falls consistent with the equation of state was $p= \epsilon$  before the
transition and frezes out again with $T \approx
.35m_f$.
However the chemical potential in this case falls  faster than the
temperature.
When one traverse  the tricritical point the distortion of the
Fermi-Dirac distribution is greater than for the second-order phase
transition
 case
as shown in Fig.\ref{fig:ntri}.

\begin{figure}
   \centering
   \epsfig{figure=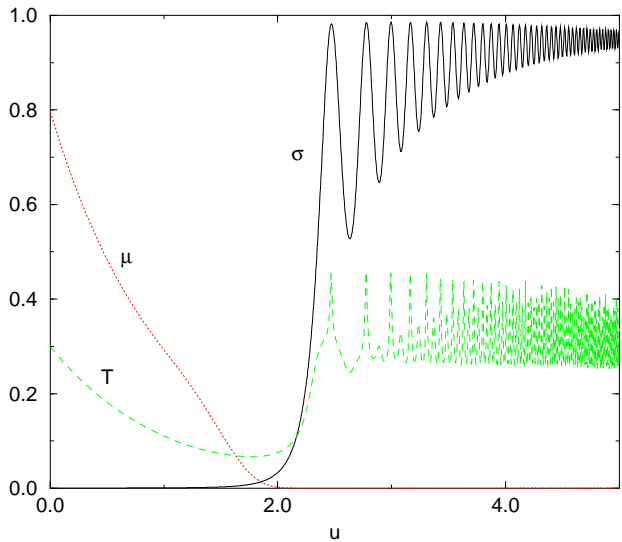,width=4.0in}
\caption{Evolution of $T$,$\mu$ and $\sigma$ as a function of $u$ passing
through first order phase transiton.}    \label{fig:firstord}
\end{figure}

\noindent {\it case (4)} $\mu_0= .8; T_0= .3$ --first order phase transition
\newline
Finally we present results when we start in the unbroken mode and go through
a first order phase transition. First we display the $\sigma, \mu, T$ in
Fig. \ref{fig:firstord}.
The results for these variables are qualitatively the same as for going
through the tricritical point, however the chemical potential falls even
faster in this case.
By comparing the three different cases where there is a phase transition,
we find  among the
parameters  $T$,$\mu$ and $\sigma$, only  the behavior of the chemical
potential $\mu$ is effected by the order
of the phase transition. The chemical potential falls as ${1 \over \tau}$
for the second order phase transition and  faster for the first order
transition suggesting a deviation from local chemical equilibrium.

\begin{figure}
   \centering
   \epsfig{figure=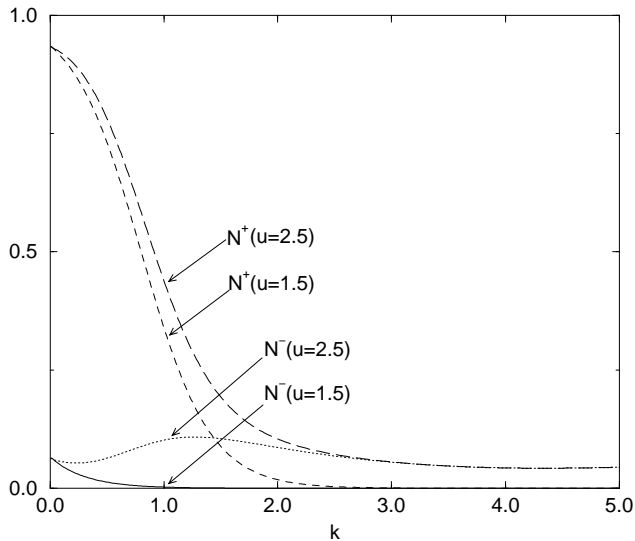,width=4.0in}
\caption{Evolution of $N_{\pm}$ as a function of $u$ when there
is a first order phase transiton. }    \label{fig:Nfirst} \end{figure}

The greatest change in the Fermi-Dirac Distribution also occurs for
the first order transition.   This
is a result of converting (the small amount of) latent heat into pair
production and is seen in Fig.
\ref{fig:Nfirst}.

As we recall from our study of the effective potential, for the GN model
the first order phase transition is not very strong.  The difference in
energy density between the false and true vacuum (measured in units of
$m_f^{-2}$) is ${1 \over 4 \pi}$, wheras the height of the barrier at phase
coexistence (seen from Fig. 2) is less than 10 \% of this difference.

\subsection {Numerical results for the energy density and pressure}

In local equilibrium  in the massless phase, we have shown that the equation
of state would be
$\epsilon=p$ and these quantities would fall as $1/\tau^2$ as shown in Fig.
\ref{fig:etherm}.  In our field theory simulations, we will find that this
behavior is followed until the  phase transition occurs.  After that the
energy density and pressure diverge from each other and oscillate. These
oscillations would be damped if we were to go beyond mean field theory and
include hard scatterings between the Fermions.  After the phase transition
we
find that the energy density oscillates around the true broken symmetry
values
discussed earlier, namely \[   \epsilon_0 = -{1 \over 4 \pi} ~. \]

\begin{figure}
   \centering
   \epsfig{figure=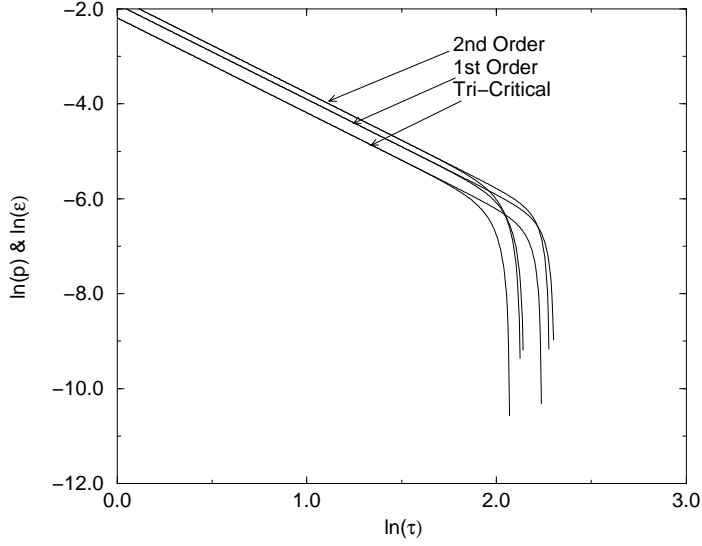,width=4.0in}
\caption{ Evolution of  $p$ and $\epsilon$ as a function of $u$ for three
initial
conditions. }
   \label{fig:epsilon}
\end{figure}

For these simulation we will assume the initial conditions described earlier
and plot the renormalized
energy density and pressure described by eq. (\ref{eq:eps}) and
 eq. (\ref{eq:p}).
Starting in
the massless (unbroken symmetry) regime  we will find $p= \epsilon$ and this
fall off pertains
until one goes through the phase transition, after which the system is no
longer in local equilibrium.
  In
Fig. \ref{fig:epsilon} we show the result of  plotting the energy density
$\epsilon(u)$  and pressure $p$ for three initial conditions.

\begin{figure}
   \centering
   \epsfig{figure=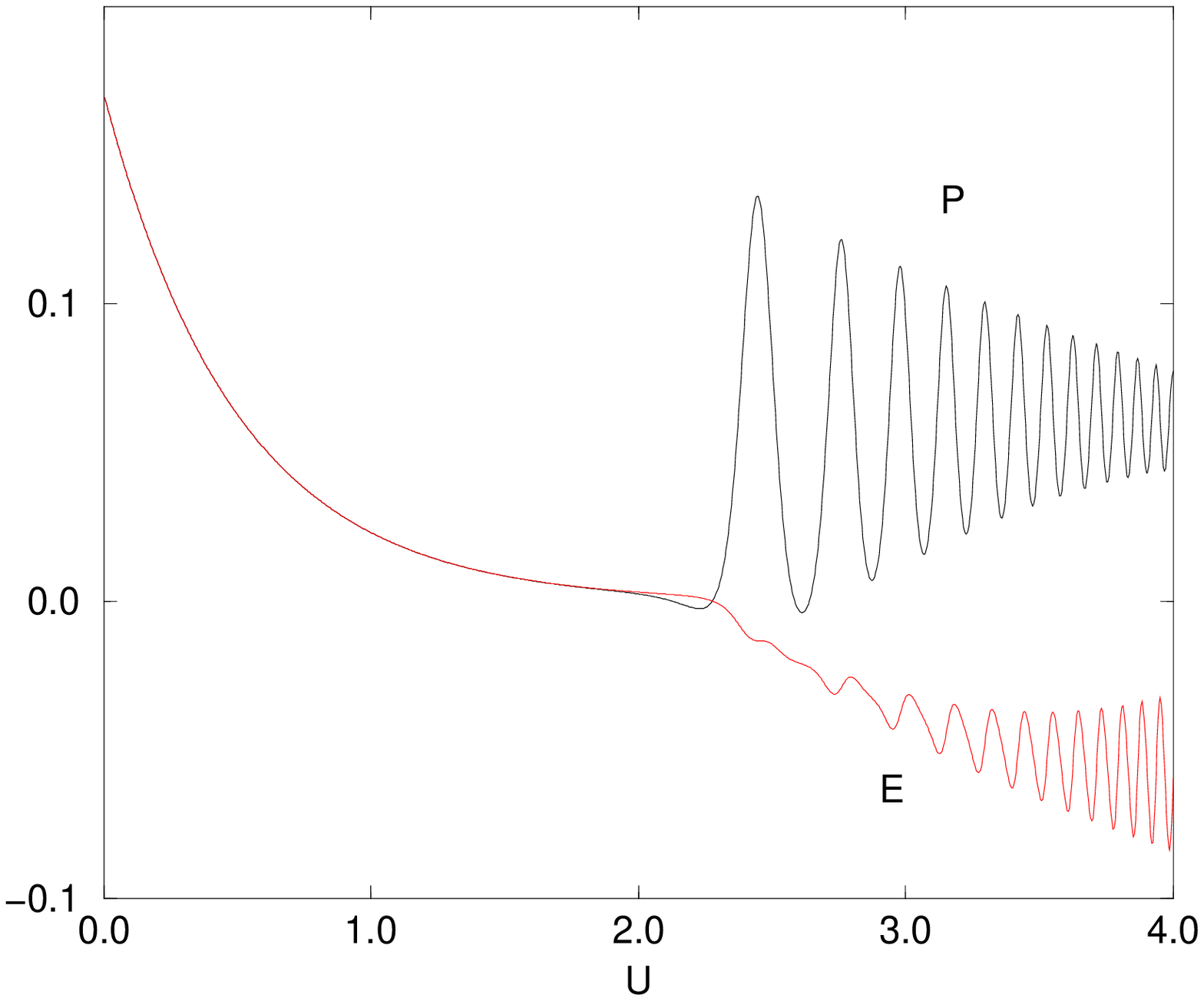,width= 2.0in,height=1.8in}
   \epsfig{figure=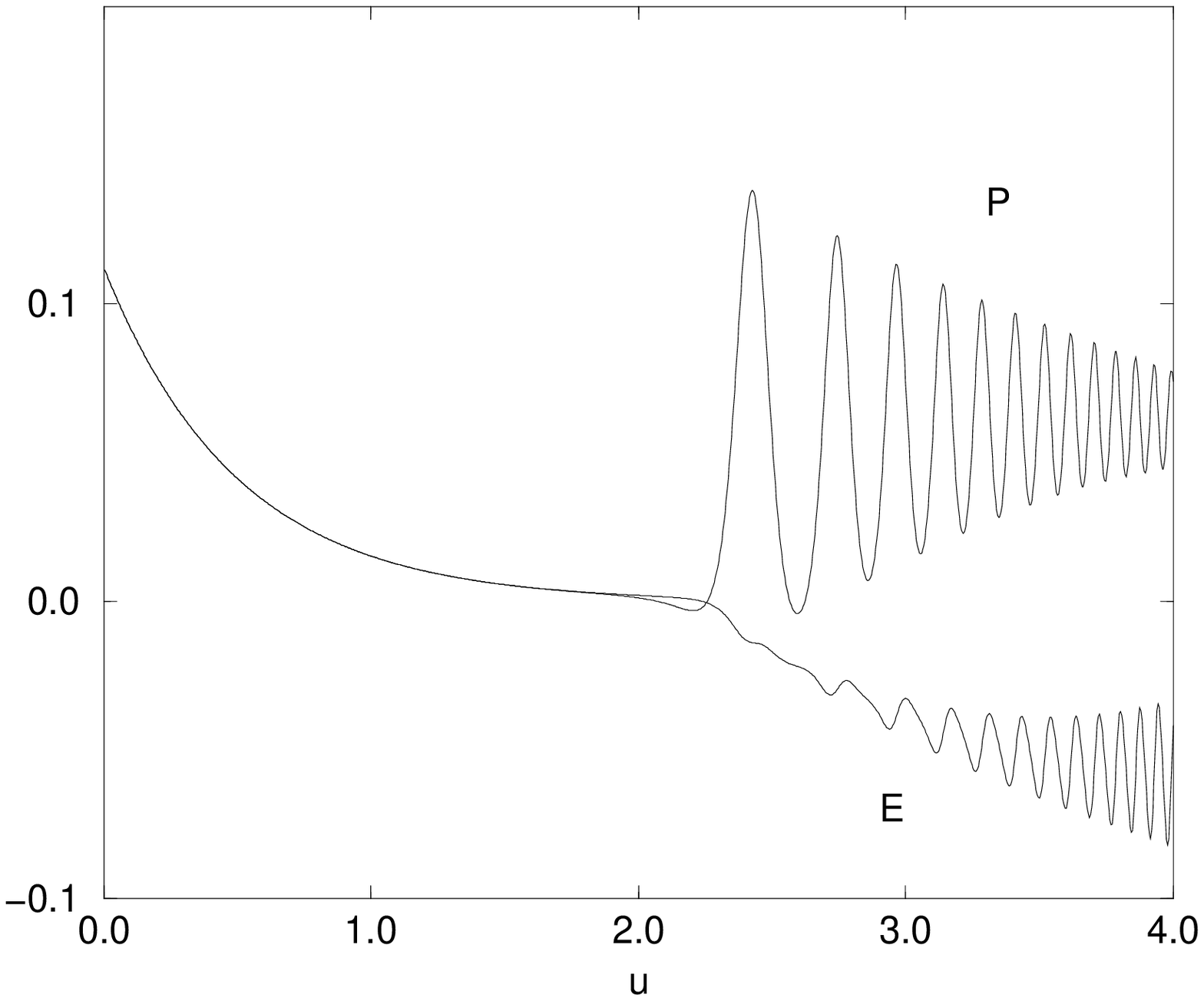,width= 2.0in,height=1.8in}
   \epsfig{figure=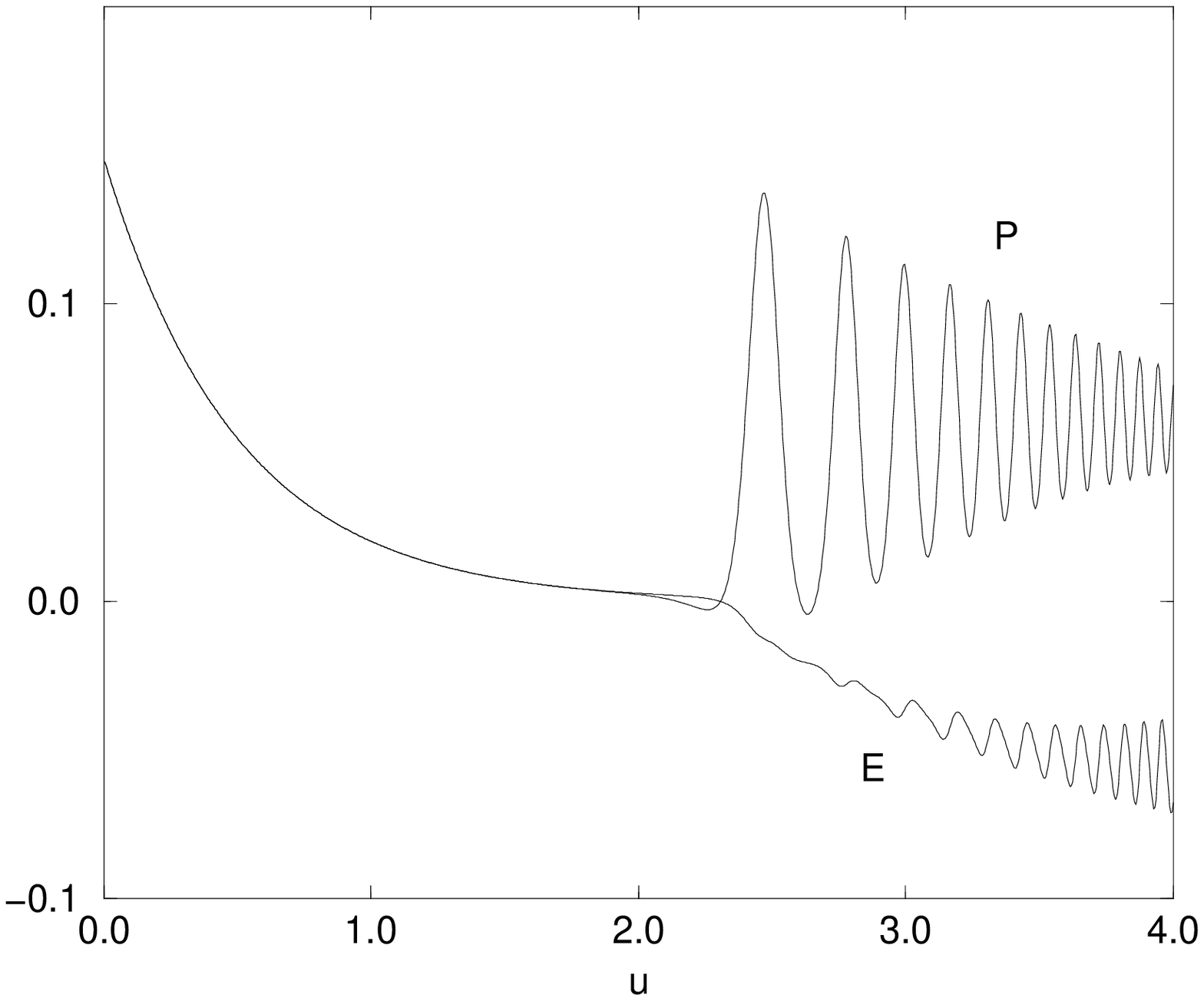,width= 2.0in,height=1.8in}
\caption{From left to right:
 evolution of the pressure  and energy density  as a function of
$u$ when there is a second order phase transition;
evolution of the pressure  and energy density  as a function of $u$ when
the trajectory passes through a tricritical point;
evolution of the pressure  and energy density  as a function of $u$ when
there is a first order phase transition.
}
   \label{fig:ep}
\end{figure}

In Fig. \ref{fig:ep} we plot  the pressure and energy density for the three
initial data
(cases (2),(3),(4) ) for which
there is a phase transition.  We see that before the phase transition, $p$
tracks $\epsilon$.
After the transition, the system goes out of equilibrium and then oscillates
about its vacuum value
because there are no damping mechanisms.

 \section{``Pion" correlation function}
One interesting question we can ask is how correlation lengths change as we
go through the phase transition.
In our toy model there is no physical pion bound state in our approximation,
but we still can define
an effective field for the pion and study the spatial dependence of the pion
green's function.
Other correlation functions one can study in a similar fashion are the
density fluctuations as well as the condensate
fluctuations.

We can define an effective (neutral) pion field via
\begin{equation}
   \pi(x) \equiv c  \bar{\psi}_i(x) i \gamma^5 \psi_i(x) ~,
\end{equation}
where $c$ is a constant.  Using our mode expansion we find that
\begin{eqnarray} \label{eq:pi}
\langle : \bar{\psi}  i \gamma^5 \psi :\rangle && =
 {2 \over \tau^2}  \langle [\phi_i(x)^{\dag}, \gamma^3 \phi_i(x) ]\rangle
\nonumber \\ &&= \int dk_{\eta} k_{\eta} [ 2- N_{+}(k) - N_{-}(k)] {d \over
du} |f_k|^2 ~.
\end{eqnarray}
Because the integrand in  eq. (\ref{eq:pi}) is odd, the expectation value
is zero (otherwise there would be spontaneous breakdown of parity).

For the equal time correlation function in LOLN, we
obtain
the usual Fermion self energy loop. This depends on the time evolving
distribution of Fermions and anti-Fermions.
 Apart from an overall  constant one can write the connected
correlation function in the form:
\begin{equation}
 D(\eta-\eta^\prime;\tau) \equiv \langle \pi(\eta,\tau)
\pi(\eta^\prime,\tau) \rangle_c  = {1 \over \tau^4}
{\rm Tr}[\gamma^3 S(\eta-\eta^\prime;\tau) \gamma^3 S(\eta^\prime -
\eta;\tau)]~,
 \end{equation}
where at equal times, the propagator is just
\begin{equation}
S(\eta,\eta^\prime;\tau)_{\alpha \beta} = \langle
[\phi_{\alpha}(\eta,\tau),\phi^{\dag}_{\beta}(\eta^\prime,\tau)] \rangle~.
\end{equation}
Here  $\alpha,\beta$ take on the values $\{1,2 \}$ and are the spinor
indices.
Using the mode expansion eq. (3.20) we find that the equal time propagator
can be
written
as
\begin{eqnarray}
S(\eta,\eta^\prime;\tau)_{\alpha \beta}&&=\int {dk_{\eta} \over 2 \pi} e^{i
k_{\eta} (\eta - \eta^\prime)}\left[ \left( 1- 2 N_+(k) \right) \phi^{+}_{k
\alpha}(\tau) \phi^{\dag +}_{k \beta}(\tau) \right. \nonumber \\
&& +  \left. \left(2 N_{-}(k) -1 \right) \phi^{}_{k
\alpha}(\tau) \phi^{\dag -}_{k \beta}(\tau) \right ]~.
\end{eqnarray}
We could also have use the mode expansion in terms of adiabatic mode
functions,
eq. (7.1) and obtained an expression for the equal time propagator in terms
of the time evoloving adiabatic number distributions.

Evaluating the trace we obtain
\begin{equation}
\tau^4  D(\eta-\eta^\prime;\tau) = \int {dk \over 2 \pi} {dq \over 2 \pi}
e^{i (k-q) (\eta-\eta^\prime)} D(k,q;\tau)  ~,
\end{equation}
where
\begin{eqnarray}
D(k,q) &&= ([1- 2 N_+(k)][1- 2 N_+(q)] + [1- 2 N_-(k)][1- 2 N_-(q)] )
F_1(k,q)
\nonumber \\ && + ([1-2N_+(k)] [1-2N_-(q)]+[1-2N_+(q)]  ~.
[1-2N_-(k)] ) F_2(k,q)
\end{eqnarray}

and
\begin{equation}
 F_1(k,q;\tau) = |f_k|^2 |f_q|^2 \{ (k \Delta_q+q \Delta_k)^2 + [
k(\tOm_q+\ts) -q(\tOm_k+\ts) ]^2 \} ~,
\end{equation}
and

\begin{eqnarray}
 F_2(k,q;\tau)&& = |f_k|^2 |f_q|^2  ( [kq -\Delta_k \Delta_q+(\tOm_k+\ts)
(\tOm_q+\ts)]^2  \nonumber \\
  && + [\Delta_k(\tOm_q+\ts) +
\Delta_q(\tOm_k+\ts)]^2  ) ~.
\end{eqnarray}

If we were solving a $3+1$ dimensional 4-fermion model with actual pion
composite particles we would now be in a position to determine
the single particle distribution function for the pions from the
Wigner distribution functions which is just a particular Fourier transform
of the Green's
function over the relative coordinate.

\begin{figure}
   \centering
    \epsfig{figure=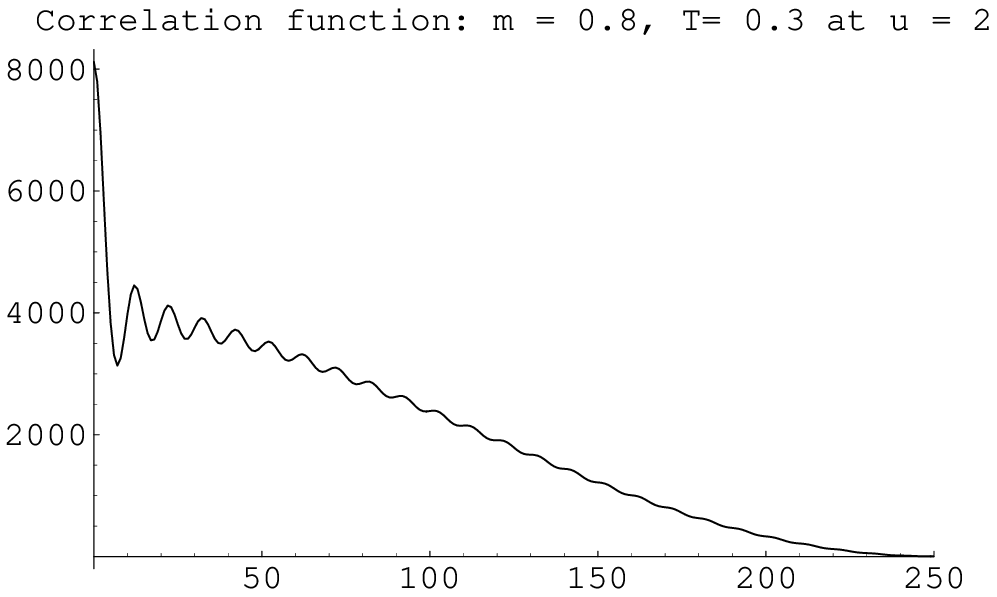,width=2.5in,height=2.0in}
    \epsfig{figure=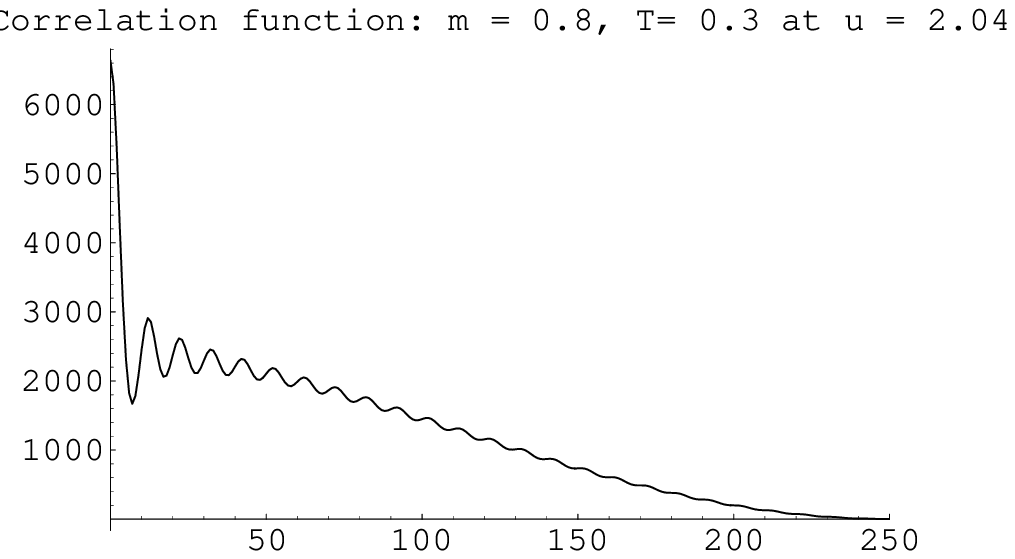,width=2.5in,height=2.0in}
    \epsfig{figure=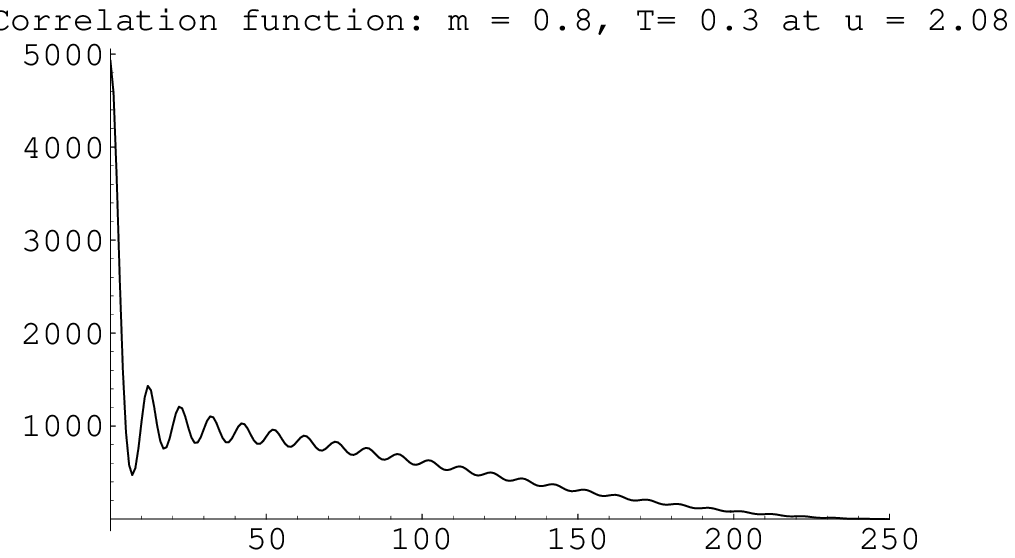,width=2.5in,height=2.0in}
    \epsfig{figure=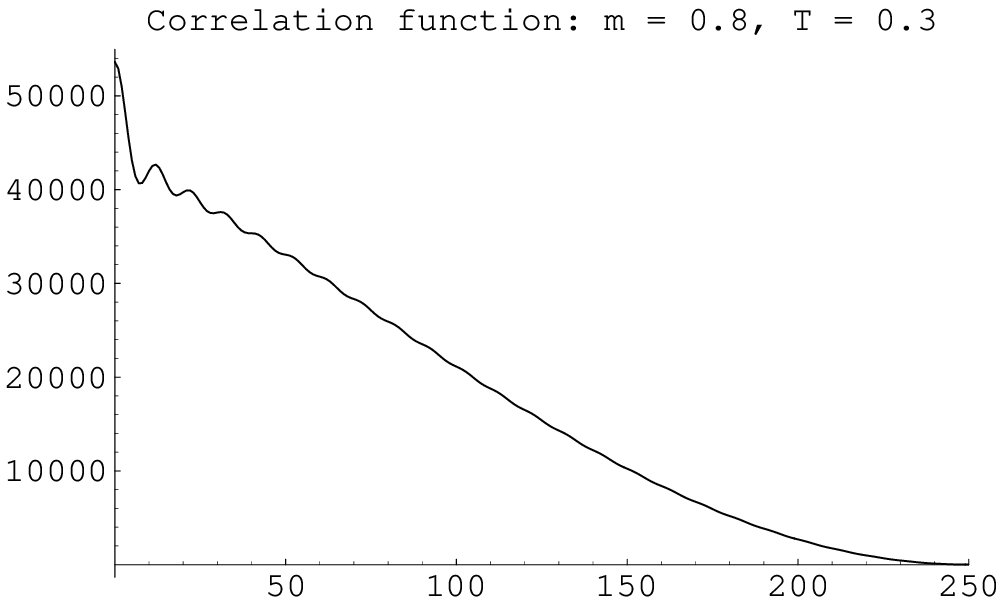,width=2.5in,height=2.0in}
\caption{Evolution of the Correlation function as a function of time.
This is for a first order transition.}
\label{fig:Vfirst} \end{figure}

\begin{figure}
   \centering
    \epsfig{figure=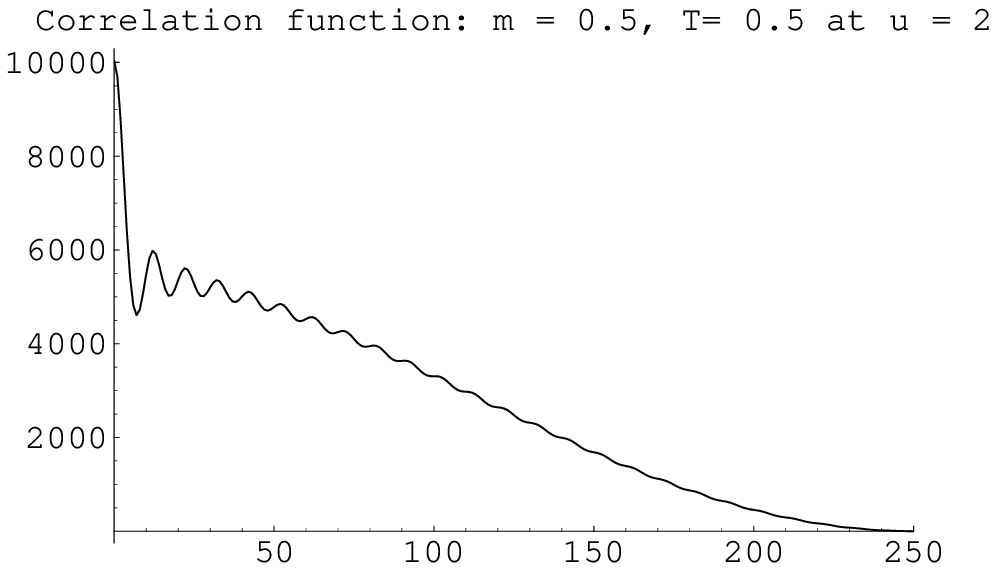,width=2.5in,height=2.0in}
    \epsfig{figure=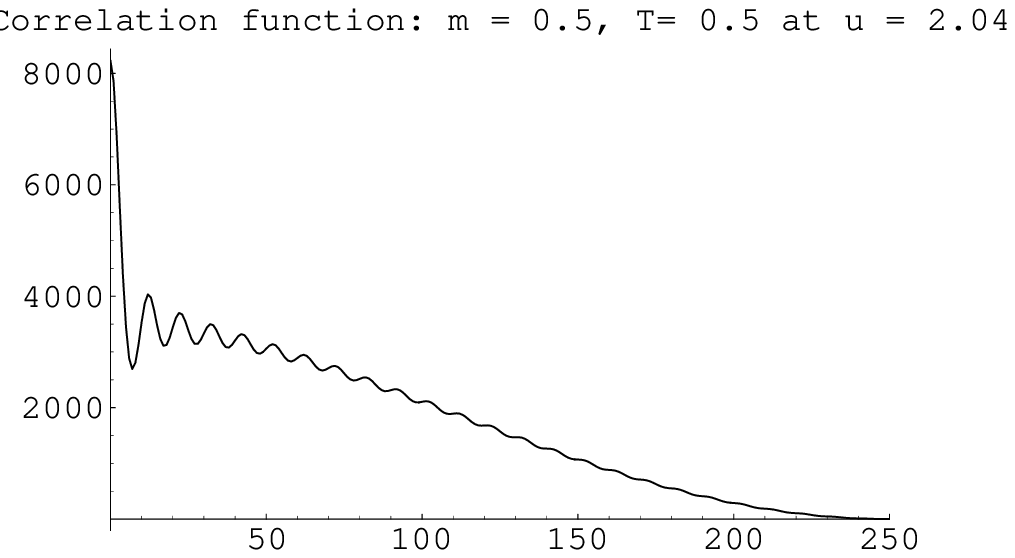,width=2.5in,height=2.0in}
    \epsfig{figure=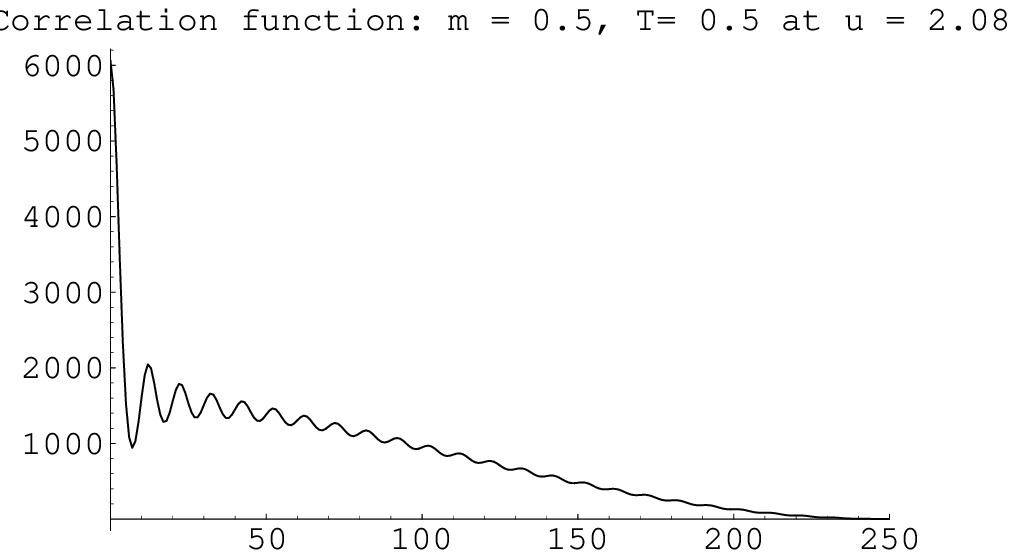,width=2.5in,height=2.0in}
    \epsfig{figure=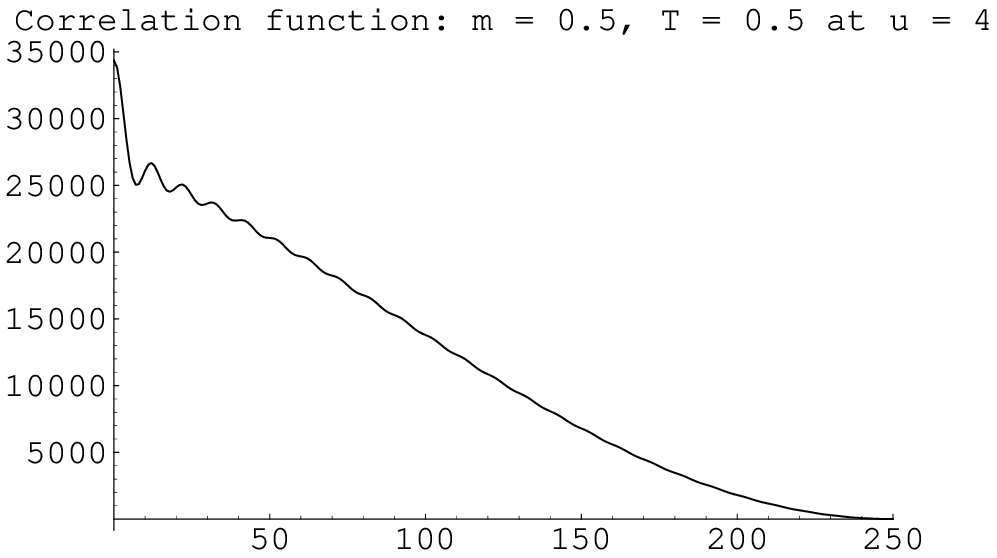,width=2.5in,height=2.0in}
\caption{Evolution of the Correlation function as a function of time.
This is for a second order transition.}
\label{fig:Vsecond} \end{figure}

 In
Figs. \ref{fig:Vfirst}, \ref{fig:Vsecond}
we plot $ \tau^4 D(\eta, \tau_f)$ as a function of $\eta$  for cases (2) and
(4) for
a sequence of times starting near the onset of the phase transition.

\section{Conclusion}
In this paper we have performed simulations in the Gross-Neveu model of an
expanding plasma of fermions and
antifermions. We have chosen inital conditions where the density matrix is
described by   single particle distribution
functions pertinent to a plasma initially in local
thermodynamic and chemical  equilibrium.   The model was treated in
the leading order in large $N$
approximation. In this approximation the  phase diagram at finite
chemical potential and temperature shares features with that of
massless 2-flavor  QCD.  We found
that if we start in the unbroken symmetry phase, the system remains in
equilibrium until the phase
transition and then goes rapidly out of equilibrium as the phase transition
is traversed.  The effects of
the phase transition are greatest when we traverse a first order phase
transition and are most noticable in
the anti-fermion distribution function. If these effects survive hard scatterings
then this should have an effect on the distribution of dileptons, just
as the overpopulation of soft pions during DCC production effected
the distruibution of dileptons as discussed by us earlier \cite{slansky}.
 We also find that before the phase transition, the system
behaves identically to an ideal fluid in local thermal equilibrium with
equation of state $p=\epsilon$. After the phase transition, the system
quickly
reaches its true broken symmetry vacuum value for the fermion mass and for
the
energy density.  Since hard scatterings are ignored in this approximation,
the competition between  the expansion of the plasma and the competing
process
for re-equilbration could not be studied here. Also by restricting
our simulations to inhomogeneities which are boost invariant, we
were not able to look at bubble nucleation.  In future
investigations we will attempt to remedy the shortcomings just mentioned.
We will perform mean field simulations for inhomogeneous initial conditions
which we discussed before \cite{inhom} and which have already been undertaken
in scalar field theory in 1+1 dimensions \cite{ref:Aarts}.  There are
also now two different approaches for going beyond $1/N$ based
on Schwinger Dyson equations which we hope we can implement to 
study  whether rethermalization  can occur for the type of expansion
expected following a heavy ion collision.  These approaches are based
on different approximations for the generating functional for the 2-PI irreducible
graphs  \cite{ref:beyond} 
\cite{ref:berges} and should allow us to study  the question of
rethermalization. We also want to extend our simulation to a more realistic
O(4) 4-Fermi Model in $3+1$ dimensions so that we can directly study
the effect of the phase transition on pion correlation functions.  

\begin{acknowledgments}
We would like to thank Emil Mottola for his help with the renormalization of
the energy momentum tensor, Salman Habib for his help with our numerical
strategy, Gordon Baym for suggesting we study the pion correlation
function and the participants of the Riken workshop on in and out of
equilibrium physics (July 200), especially
Dan Boyanovsky for helping us clarify some issues.

\end{acknowledgments}

%




\newcounter{mysec}
\newcommand{\myappendix}{\appendix
\setcounter{mysec}{0}
 \renewcommand{\themysec}{\Alph{mysec}}}
\newcommand{\myappsection}[1]{
 \setcounter{equation}{0}
  \addtocounter{mysec}{1}
\section*{\themysec\ \ #1}}
\renewcommand{\theequation}{\themysec.\arabic{equation}}


\myappendix

\myappsection{spinor bases, Dirac matrices}

We will choose for convenience the following representation
for the matrices $\gamma^0$ and $ \gamma^3$
\begin{equation}
 i \gamma^0 =   \left( \matrix{ 1  &  0 \cr
0  & -1  \cr}  \right)
\end{equation}

\[
\gamma^3 =  \left( \matrix{ 0 & 1 \cr
1 & 0 \cr} \right)
\]
Thus the spinor eigenstates of $ i \gamma^0$ are
\[
\chi^+ =   \left[ \matrix{  1  \cr
0 \cr} \right]
\]

\[
\chi^- =  \left[ \matrix{ 0 \cr
1 \cr} \right]
\]

In terms of this explicit representation we find that the momentum space
wave functions have the form:
\begin{eqnarray}
\phi^+_k && = \left[ (\tOm_k+ \ts -i \Delta_k) \chi^{+} - i k_{\eta}
\chi^{-}  \right] f^+_k \nonumber \\
\phi^-_k && =\left[ (\tOm_k+ \ts+ i \Delta_k) \chi^{-} - i k_{\eta} \chi^{+}
\right] f^{-}_k
\end{eqnarray}

\end{document}